\documentclass[reprint, amsmath, amssymb, aps]{revtex4-2}

\usepackage{comment}
\usepackage{graphicx}
\usepackage{bbold}
\usepackage{xcolor}
\usepackage{dcolumn}
\usepackage{bm}
\usepackage{bbm}
\usepackage{braket}
\usepackage{mathdots}
\usepackage[caption=false]{subfig}
\usepackage{overpic}
\usepackage{hyperref}
\usepackage{multirow}
\usepackage{amssymb}
\usepackage{pifont}
\usepackage{bbding}

\newcommand{\munit}[1]{m_{#1}}

\newcommand{\calP}{{\mathcal{P}}}
\newcommand{\calH}{{\mathcal{H}}}
\newcommand{\calV}{{\mathcal{V}}}
\newcommand{\calL}{{\mathcal{L}}}
\newcommand{\calQ}{{\mathcal{Q}}}
\newcommand{\calD}{{\mathcal{D}}}

\newcommand{\notzerop}{\#}
\newcommand{\cmark}{\ding{51}}
\newcommand{\xmark}{\ding{55}}

\begin{document}

\title{Motion from Measurement:\\ The Role of Symmetry of Quantum Measurements}

\author{Luka Antonić}
\author{Yariv Kafri}
\author{Daniel Podolsky}
 \email{podolsky@physics.technion.ac.il}
\author{Ari M. Turner}
\affiliation{
Department of Physics, Technion -- Israel Institute of Technology, Haifa 3200003, Israel
}

\date{\today}

\begin{abstract}
In quantum mechanics, measurements are dynamical processes and thus they should be capable of inducing currents.  The symmetries of the Hamiltonian and measurement operator provide an organizing principle for understanding the conditions for such currents to emerge.  The central role is played by the inversion and time-reversal symmetries.  We classify the distinct behaviors that emerge from single and repeated measurements, with and without coupling to a dissipative bath. While the breaking of inversion symmetry alone is sufficient to generate currents through measurements, the breaking of time-reversal symmetry by the measurement operator leads to a dramatic increase in the magnitude of the currents. We consider the dependence on the measurement rate and find that the current is non-monotonic.  Furthermore, nondegenerate measurements can lead to current loops within the steady state even in the Zeno limit.
\end{abstract}

\maketitle

\section{Introduction}
\label{sec:Intro}

One of the most exciting properties of quantum mechanics is the ability to influence the dynamics and steady-state properties of systems through measurement.  Notable examples include the Zeno effect, where the dynamics of a system under constant observation can be stalled or severely restricted~\cite{misra1977, gourgy2018, blumenthal2022}, the ability to use quantum measurements to entangle qubits~\cite{raussendorf2001, nielsen2006, burgath2014, gershoni2023, bairey2019}, control quantum computations~\cite{tantivasadakarn2022, verresen2023} and dynamics~\cite{grunbaum2013, yin2024, dhar2015, friedman2017, popperl2023, popperl2024}, and induce currents\cite{wampler2022, khor2023, ferreira2024}.  More recently, the remarkable possibility that measurements in quantum circuits can induce phase transitions in the nature of the entanglement has been pointed out~\cite{li2018, skinner2019, chan2019, buchold2021, noel2022, ladewig2022, poboiko2023, poboiko2024}.  

Much of the above body of work relies on quantum measurements disrupting the unitary evolution.  However, a salient feature which is typically ignored is the symmetries of the measurement operator and their interplay with those of the Hamiltonian.  In this paper, we explore the ramifications of symmetries for the question ``Can quantum measurements induce currents, and if so, how?'' For instance, it is interesting to ask whether a series of measurements of observables with time-reversal symmetry can create currents due to the irreversibility of the measurements, or if explicit time-reversal symmetry breaking is necessary.  Throughout the paper, we focus on local projective measurements and consider the current resulting from averaging over an ensemble of such experiments without post-selection. 

To address this, we begin by identifying two distinct sources of charge transfer.  The first is the standard current operator, which we dub the {\em Hamiltonian current}.  The second is the {\em measurement charge displacement}, which occurs at the instant of measurement and is independent of the Hamiltonian.  

Considering first a single measurement applied to a system in equilibrium, we find that a current can be produced only if inversion is broken, either in the Hamiltonian $H$ or the observable $A$ (for a measurement, a symmetry $S$ is preserved if and only if $SAS^{-1}=\pm A$).  Given that inversion symmetry is broken, if time reversal is preserved then a current can be generated but it either oscillates or decays after the measurement.  In sharp contrast, measurements that break time-reversal symmetry lead to a {\em DC current}~\footnote{A Hamiltonian that breaks time reversal and inversion generically leads to currents, even in equilibrium, at least for finite systems (Bloch's theorem states that these currents vanish in the thermodynamic limit \cite{watanabe_2019}).  We therefore restrict our attention to time-reversal symmetric Hamiltonians that may or may not have an inversion symmetry.}.

We then turn to investigate an isolated system subjected to repeated measurements.  In this case, the steady state is an infinite temperature state, which we show not to support currents or measurement charge displacements. This fate is avoided by coupling the system to a thermal bath, which provides dissipation and leads to a non-trivial steady state.  Then, currents develop provided that inversion symmetry is broken, and their magnitude is a non-monotonic function of the measurement rate.  Also here the time reversal properties of $A$ play a central role. If the measurement preserves time-reversal symmetry then the current requires strong dissipation to be sizable. By contrast, if $A$ breaks time reversal, a large steady-state current develops, even for weak dissipation.

In the particular case where the measurement preserves inversion, but the Hamiltonian does not, by analogy with classical \cite{smoluchowski1927, feynman_leighton_sands_1963, brownian_review, magnasco1993, peskin1993, rousselet1994,julicher1997} and quantum \cite{reimann, reimann1997, scheidl,salger2009} ratchets, we describe the currents as resulting from a quantum-measurement ratchet effect. This is a novel type of ratchet effect, where an inversion-preserving measurement, which acts as an unbiased drive, leads nonetheless to a directed current.  

Time-reversal symmetry also plays an important role in the Zeno limit. It is often stated that a watched pot never boils~\cite{peres1980} to refer to the fact that non-degenerate measurements lead to frozen dynamics.  Here, we interestingly find that local current loops develop when the measurement breaks time-reversal symmetry.  In other words, a watched pot can convect.

Our analysis considers both stochastic Poisson measurements and periodic ones.  In the latter case, we find resonances when the measurement period matches the natural dynamical scales of the Hamiltonian.

\begin{figure}
\includegraphics[width=.48\textwidth]{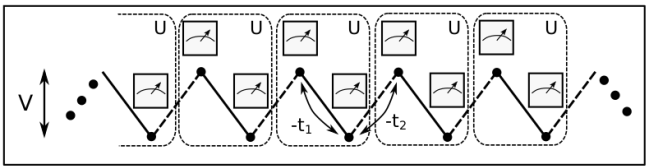}
\caption{One-dimensional chain with two sites in a unit cell, with alternating hopping amplitudes $t_1$ and $t_2$, and with a staggered potential $V$, which breaks inversion. A measurement consists of acting with a $2\times 2$ unitary $U$ on all cells simultaneously, measuring the location of the particle, and then acting with the inverse transformation $U^\dagger$.}
    \label{fig:measurement_protocol}
\end{figure}

To make the discussion concrete, we restrict our discussion to one-dimensional lattice systems and demonstrate our results on a specific example -- a single spinless particle hopping on a dimerized chain with amplitudes $t_1\ne t_2$ and staggered potential $V$, with the Rice-Mele Hamiltonian (see Fig.~\ref{fig:measurement_protocol})
\begin{equation}\label{eq:hamiltonian}
\begin{split}
    H =& \sum_{n=1}^N\Big[-t_1c_{2n-1}^{\dagger}c_{2n}-t_2c_{2n}^{\dagger}c_{2n+1}+\mathrm{h.c.}\\ &\ \ \ \ +\frac{V}{2}\left(c_{2n-1}^{\dagger}c_{2n-1}-c_{2n}^{\dagger}c_{2n}\right)\Big]\,.
\end{split}
\end{equation}
Here, $N$ is the number of unit cells in the chain, and we assume periodic boundary conditions $c_{n+2N}^{\dagger}=c_{n}^\dagger$.  Inversion symmetry is broken provided $V\ne 0$. The eigenstates of the Hamiltonian are fully specified by the wavevector $k$ and band-index $\mu\in\{+,-\}$ with energies 
$E_{k, \mu}=\mu\sqrt{t_1^2+t_2^2+2t_1t_2\cos(2k)+\frac{V^2}{4}}$, where $k=\frac{n\pi}{N}, n\in\{0,1,...,N-1\}$.  When appropriate, we will also consider coupling to a thermal bath. 

We consider a measurement that determines the unit cell that the particle occupies as well as the value of an operator within the unit cell.  This can be written as a single measurement of an operator,
\begin{equation}\label{eq:observable}
A=\bigoplus_{n=1}^{N}\left(2n+\frac{\mathbf{\hat{m}}\cdot\boldsymbol{\sigma}-1}{2}\right),
\end{equation}
where $\mathbf{\hat{m}}$ is a unit vector that determines the specific measurement and $\sigma_{x,y,z}$ are the Pauli matrices acting on the 2-dimensional Hilbert space within a unit cell. 
Note that for $\mathbf{\hat{m}}=\mathbf{\hat{e}}_z$, $A$ corresponds to a measurement of the exact location of the particle, where $2n$ distinguishes between different unit cells and $(\sigma_z-1)/2$ specifies the site within the unit cell.  For general $\mathbf{\hat{m}}$, $A$ can be implemented in practice through an associated $2\times 2$ unitary transformation $U$ which satisfies $U^\dagger\sigma_zU=\mathbf{\hat{m}}\cdot\boldsymbol{\sigma}$.  The unitary $U$ is applied simultaneously on all the unit cells, as sketched in Fig. \ref{fig:measurement_protocol}.  The measurement of the observable $A$ is then a composite operation, consisting of the unitary $U$, a measurement of location, and the inverse unitary $U^\dagger$, performed in rapid succession.

The paper is organized as follows.  In Sec.~\ref{sec:definitions} we discuss different kinds of charge displacements that can occur in a quantum system with measurements.  In Sec.~\ref{sec:symmetry} we use symmetry arguments to understand the characteristics of the currents that occur when a measurement is performed starting in thermal equilibrium.  In Sec.~\ref{sec:InfiniteT} we consider the steady state resulting from repeated measurements on a thermally isolated system.  In Sec.~\ref{sec:thermalBath} we couple the system to a thermal bath using the Lindblad formalism and study the case of random Poisson measurements. Finally, in Sec.~\ref{sec:Periodic} we study the complementary case of Floquet measurements.

\section{Charge displacement in a quantum system with measurements}\label{sec:definitions}

The current operator on a lattice can be defined from the continuity equation, $\dot{n}(x)=j(x-1)-j(x)$, where $n(x)$ is the density at site $x$ and $j(x)$ is the current from $x$ to $x+1$.  The center-of-mass velocity is equal to the time derivative of the position operator $\hat{x}=\sum_x x\, n(x)$.  This is given by 
\begin{align}
\frac{d\hat{x}}{dt}&=\sum_x x\, \dot{n}(x) = \sum_x x\left[j(x-1)-j(x)\right]\\
&=\sum_x j(x)=J\,,
\end{align}
where $J\equiv\sum_x j(x)$ is the current operator summed over all space. The average charge displacement in the time interval $(t_i,t_f)$ is thus
\begin{align}\label{eq:hamiltonian_charge_displacement}
\Delta x=\int_{t_i}^{t_f}dt\,  \langle {J} \rangle\,.
\end{align}
Here angular brackets denote an average with respect to the density matrix.

In a {\em closed quantum system} with unitary evolution governed by the Hamiltonian $H$, the velocity operator is given by the Heisenberg equation of motion for $\hat{x}$.  Thus,
\begin{align}\label{eq:J_H}
J_H=-i\left[H,\hat{x}\right]\,,
\end{align}
where we have added the subscript $H$ to indicate that this is the current resulting from the unitary dynamics.  We call this type of current the {\em Hamiltonian current}.

When quantum measurements are added to the mix, the Hamiltonian current defined above is not enough to capture all particle motion~\cite{avron2012, hovhannisyan2019}.  A projective measurement of an observable $A$ leads to a sudden change in the density matrix
\begin{equation}\label{eq:density_matrix_mapping}
    \rho \mapsto\rho'=\sum_{a}P_a\rho P_a\,,
\end{equation}
where $P_a$ is the operator that projects to the eigenspace of $A$ with eigenvalue $a$.   The sudden nature of projective measurements renders the definition of a current unnatural.  Therefore we consider the average displacement of the position of the particle as a result of this measurement, which is given by
\begin{align}
    \Delta {  x}_A  = \mathrm{tr}\left[(\rho' - \rho) \hat{{  x}}\right] = \mathrm{tr}\left[\rho\left(\sum_a P_a\hat{{  x}}P_a-\hat{{  x}}\right)\right]\,.
\end{align}
This is the expectation value of the operator
\begin{align}
    \hat{  Q}=\sum_a P_a\hat{{x}}P_a-\hat{{  x}} \label{eq:Qx}
\end{align}
in the pre-measurement state $\rho$.  We call $\hat{  Q}$ the {\em measurement charge displacement}.  It describes a charge transfer that is sudden, to the extent that the measurement is instantaneous.  In a system with unitary Hamiltonian dynamics interspersed with measurements, the total charge motion can be obtained from the integral of the Hamiltonian current over the periods of unitary evolution, together with the charge displacement at the instants of measurement.

In systems with periodic boundary conditions, the position operator $\hat{x}$ is ill-defined, rendering the definition of the measurement charge displacement operator problematic.  To sidestep this issue, $\hat{Q}$ can be rewritten as
\begin{align}
\hat{Q}=\frac{1}{2}\sum_a \left(P_a \left[\hat{x},P_a\right]- \left[\hat{x},P_a\right] P_a\right)\,.
\end{align}
This involves only the commutator of $\hat{x}$ with a local operator, which is well-defined.

Note that the definition of $\hat{Q}$ depends only on the choice of measurement operator $A$, and is independent of the Hamiltonian.  In the special case where the operator $A$ is the position itself, $\hat{Q}$ vanishes exactly, since in that case $P_a$ commutes with $\hat{x}$ in Eq.~\eqref{eq:Qx}.

\section{Symmetry considerations for a single measurement}\label{sec:symmetry}

In this section, we will consider the creation of currents by a single measurement starting from equilibrium.  We will discuss the restrictions that discrete symmetries place on the charge displacements and currents discussed in the previous section.  We consider three symmetries: spatial inversion $I$, time reversal $T$, and their product $IT$ (more commonly referred to as PT symmetry).  The results will play an important role later, when we discuss multiple measurements.

\subsection{Symmetry transformations of the current and displacement operators}

Let us first recall the symmetry properties of the Hamiltonian current $J_H=-i[H,\hat{x}]$. This current is odd under time-reversal symmetry $T$ and inversion $I$, provided that $H$ is invariant under these symmetries.    
\begin{equation}\label{eq:j_conditions}
    SHS^{-1}=H \Rightarrow \begin{cases}
        SJ_{H}S^{-1}=-J_{H}, \; &S\in\{I, T\},\\
        SJ_{H}S^{-1}=J_{H},\; &S=IT.
    \end{cases}
\end{equation}
Note that if $H$ is not invariant under these symmetries, then $J_{H}$ will in general not have simple transformation rules.

We now turn to the measurement charge displacement $\hat{Q}$.  For this, we must consider the transformation properties of the measured observable $A$ under the symmetry operations.  Using a common symbol $S\in\{I, T, IT\}$, a special distinction occurs for an observable $A$ that has well-defined parity (either even or odd) under the action of $S$.
For such observables, we will show that $\hat{Q}$ transforms simply under $S$,
\begin{equation}\label{eq:Q_conditions}
 SAS^{-1}=\pm A\Rightarrow     \begin{cases}
        S\hat{Q}S^{-1}=-\hat{Q}, \; &S\in\{I, IT\},\\
        S\hat{Q}S^{-1}=\hat{Q},\; &S=T.
    \end{cases}
\end{equation}
so that $\hat{Q}$ is even under time-reversal and odd under inversion and $IT$ symmetries. On the other hand, if $A$ does not have well-defined parity under these symmetries, then we cannot make any general claim regarding the symmetry properties of the measurement charge displacement operator.

To demonstrate Eq.~\eqref{eq:Q_conditions} we note that, for an eigenstate $\ket{a}$ of $A$ with eigenvalue $a$, the symmetry-related eigenstate $S\ket{a}$ satisfies
\begin{equation}
AS\ket{a}=\pm SA\ket{a}=\pm a S\ket{a}, 
\label{eq:ASpsi}
\end{equation}
so that it belongs either to the same eigenspace with eigenvalue $a$ or the eigenspace with the complementary eigenvalue $-a$. Equation~\eqref{eq:ASpsi} implies that the symmetry operator maps projectors onto projectors,~\footnote{Considering that time reversal is anti-unitary, Eq.~\eqref{eq:sasi} requires attention. If we write $P_{a}=\sum\limits_{\lambda}\ket{a,\lambda}\bra{a, \lambda}$, where $\lambda$ is an index that labels states spanning $a$-eigenspace, and $\ket{\bar{a}}=T\ket{a}$, we need to show that $TP_{a}T^{-1}$ equals $P_{\bar{a}}=\sum\limits_{\lambda}\ket{\bar{a},\lambda}\bra{\bar{a}, \lambda}$. A generic state can be written in the eigenbasis of time-reversed states, $\ket{\Psi}=\sum_{\lambda'}c_{\bar{a},\lambda'}\ket{\bar{a},\lambda'}+\ket{\Psi^{\perp}}$.  Then,  $TP_aT^{-1}\ket{\Psi}=\sum_{\lambda,\lambda'}T\ket{a,\lambda}\bra{a,\lambda}T^{-1}c_{\bar{a},\lambda'}\ket{\bar{a},\lambda'}=\sum_{\lambda,\lambda'}c_{\bar{a},\lambda'}\ket{\bar{a},\lambda'}\braket{a,\lambda|a,\lambda'}=\sum_{\lambda}c_{\bar{a},\lambda}\ket{\bar{a},\lambda}$, which is seen to equal $P_{\bar{a}}\ket{\Psi}=\sum_{\lambda}c_{\bar{a},\lambda}\ket{\bar{a},\lambda}$.} 
\begin{equation}
    SAS^{-1}= \pm A\, \Rightarrow\, SP_{a}S^{-1}=P_{\pm a}\,.
    \label{eq:sasi}
\end{equation}
Hence, for an $S$-even observable, the projector $P_a$ is invariant under the symmetry transformation.  For an $S$-odd observable, the projector $P_0$ onto the subspace with zero eigenvalue $a=0$ is invariant, whereas the projectors onto nonzero eigenvalues $a\ne 0$ are mapped onto the complementary projectors $P_{-a}$.  

Since the definition of $\hat{Q}$, Eq.~\eqref{eq:Qx}, involves a sum over all projectors $P_a$, this permutation of projectors leaves $\hat{Q}$ unchanged.  Hence, the transformation properties of $\hat{Q}$ are determined directly by those of $\hat{x}$, which is odd under inversion and $IT$ symmetries and even under time reversal.  This implies Eq.~\eqref{eq:Q_conditions}.

We comment that for the Rice-Mele model, the observable $A$ as defined in Eq.~\eqref{eq:observable} does not have well-defined parity but can be regarded as a composite operator combining a measurement of the unit cell $n$ and the observable $\hat{\mathbf{m}}\cdot\boldsymbol{\sigma}$ within the unit cell.  Hence, the projectors $P_{a}$ are products of projectors in the two subspaces $P_a=P_{n}P_{\sigma}$.   The two parts can transform differently under the action of symmetry.  The unit cell projectors always satisfy $IP_n I^{-1}=P_{-n}$ and $TP_nT^{-1}=P_n$.  Assuming that $S(\hat{\mathbf{m}}\cdot{\boldsymbol{\sigma}})S^{-1}=\pm \hat{\mathbf{m}}\cdot{\boldsymbol{\sigma}}$, the projectors $P_{\sigma}$ are either left invariant or permuted according to Eq.~(\ref{eq:sasi}).  Hence, $P_a=P_{n}P_{\sigma}$ are permuted by the action of the symmetries too, and all the reasoning still holds.  Therefore, it is enough to consider the symmetry of $\hat{\mathbf{m}}\cdot{\boldsymbol{\sigma}}$. In what follows we will write $SAS^{-1}=\pm A$ as a shorthand for
$S(\hat{\mathbf{m}}\cdot{\boldsymbol{\sigma}})S^{-1}=\pm \hat{\mathbf{m}}\cdot{\boldsymbol{\sigma}}$.

With the above insights we now turn to discuss the role of the symmetries in measurement-induced currents.  

\subsection{Currents require breaking of inversion symmetry}
\label{sec:inversion}

Inversion symmetry plays an important role in restricting currents.  Consider a Hamiltonian that is inversion symmetric, $IHI^{-1}=H$, and suppose that the observable $A$ has well-defined parity under inversion, $IAI^{-1}=\pm A$. If the system starts in thermal equilibrium, then no currents or measurement charge displacements will develop after a measurement of $A$, even if multiple measurements of $A$ are carried out in succession at different times.

To see this, note that the equilibrium density matrix $\rho_{\mathrm{eq}}=\frac{e^{-\beta H}}{Z}$ at temperature $T=1/\beta$ is invariant under inversion, $I \rho_{\mathrm{eq}} I^{-1}=\rho_{\mathrm{eq}}$.  The state immediately following a projective measurement of $A$ is
\begin{equation}\label{eq:post_measurement_state}
    \rho_0 = \sum_{a}P_{a}\rho_{\mathrm{eq}}P_{a}.
\end{equation}
If $IAI^{-1}=\pm A$ then $I$ leads to a permutation of projectors, see Eq.~\eqref{eq:sasi}, which leaves $\rho_0$ invariant.  Furthermore, unitary evolution by $H$ preserves the inversion symmetry.  We conclude that $I\rho(t) I^{-1}=\rho(t)$ at all times.  Since the Hamiltonian current is odd under inversion, Eq.~\eqref{eq:j_conditions}, this implies that
\begin{equation}\label{eq:current_expectationEq}
 \begin{split}\braket{J_H}(t)&=\mathrm{tr}\left(J_H\rho(t)\right)=\mathrm{tr}\left(IJ_HI^{-1}I\rho(t) I^{-1}\right)\\&=-\mathrm{tr}(J_H\rho(t))=0,
\end{split}
\end{equation}
{\em i.e.} the Hamiltonian current vanishes at all times.  Similarly, the measurement charge displacement $\hat{Q}$ is also odd under inversion, see Eq.~\eqref{eq:Q_conditions}, and hence it also vanishes for all measurements.

In the specific case where the observable is non-degenerate, there is a simple intuition behind this result. Inversion-even observables collapse the system to states that do not carry a Hamiltonian current \footnote{This argument relies on the observable being non-degenerate.  However, the proof is more general.}.  Inversion-odd observables, on the other hand, collapse the system to current-carrying states.  However, these states come in complementary pairs that carry opposite currents and that are equally likely to occur in thermal equilibrium, leading to a vanishing current expectation value.

This result implies that the generation of currents requires the breaking of inversion symmetry, either by $A$ or $H$.  For the former, we mean that $A$ does not have a well-defined parity under inversion.  In the next subsection, we will assume that inversion is broken, either by $A$ or $H$.  Finally, we note that the conclusions are unchanged even for initial states that are not thermal, as long as they are invariant under inversion.

\subsection{DC currents require breaking of time-reversal symmetry}
\label{sec:timeReversalBreaking}

Since currents require time-reversal symmetry breaking, one would naively think that $T$ plays an equivalent role to inversion symmetry in preventing currents. In particular, one might expect currents to vanish if $THT^{-1}=H$ and $TAT^{-1}=\pm A$.  However, as we show now, time reversal allows currents, but constrains their time dependence.

We will first show that time reversal implies the vanishing of the current immediately after a measurement starting from an equilibrium state. Recall that $THT^{-1}=H$ and assume that $TAT^{-1}=\pm A$.  Then, the equilibrium density matrix is time-reversal invariant, $T\rho_{\mathrm{eq}} T^{-1}=\rho_{\mathrm{eq}}$.  Immediately after a measurement, the density matrix remains time-reversal invariant, since according to Eq.~\eqref{eq:sasi} $T$ only permutes the projectors in the sum in Eq.~\eqref{eq:post_measurement_state}.  This implies that the Hamiltonian current {\em immediately} post-measurement vanishes.  The intuition for this result is similar to that leading to the lack of currents for observables with a well-defined inversion parity.   This would not be the case if $TAT^{-1}\ne \pm A$.

However, the current does not remain zero because the evolution of the state after the measurement breaks time reversal, as will be seen.  This produces currents, but we now show that they are either oscillating or decaying in nature, so that the net displacement after a long time is finite.  To see this, we write the density matrix as
\begin{eqnarray}
\rho(t)=\sum_{k,\alpha,\beta} c_{k}^{\alpha\beta}(t)  \ket{k,\alpha}\bra{k,\beta}
\end{eqnarray}
where $\ket{k, \alpha}$ is an eigenstate of the Hamiltonian with energy $\epsilon_{k, \alpha}$.  Here $k$ is a crystal momentum and $\alpha$ is a band index arising from orbital degrees of freedom so that it is not changed upon time reversal.  Here, we have assumed translational symmetry, which ensures that $\rho(t)$ is diagonal in $k$.  Immediately following a measurement at $t=0$, $T\rho(0) T^{-1}=\rho(0)$ implies that the coefficients satisfy 
\begin{equation}
\label{eq:ckTRS}
c_{k}^{\alpha\beta}(0)=\left(c_{-k}^{\alpha\beta}(0)\right)^*=c_{-k}^{\beta\alpha}(0)    
\end{equation}
where the second equality follows from Hermiticity of $\rho$. This condition represents time-reversal symmetry for the density matrix at $t=0$.  Time evolution leads to $c_{k}^{\alpha\beta}(t)=c_{k}^{\alpha\beta}(0)e^{-i \omega_{k,\alpha\beta} t}$, where $\omega_{k,\alpha\beta}=\epsilon_{k,\alpha}-\epsilon_{k,\beta}$, which disrupts this condition.

The Hamiltonian current $\braket{J_H}(t)=\mathrm{tr}\left[\rho(t)J_{H}\right]$ is then
\begin{equation}
\label{eq:current_at_t}
 \braket{J_H}(t)=\sum_{k,\alpha, \beta}c_{k}^{\alpha\beta}(0)e^{-i\omega_{k,\alpha\beta} t}\bra{k,\beta}
 J_H\ket{k,\alpha}
\end{equation}
To proceed, we distinguish between intraband ($\alpha=\beta$) and interband ($\alpha\ne\beta$) terms in Eq.~\eqref{eq:current_at_t}.  The intraband terms are time-independent since $\omega_{k,\alpha\alpha}=0$.  They equal
\begin{equation}
 \braket{J_H}_{\mathrm{intra}}=\frac{1}{2}\sum_{k, \alpha}\left[c_{k}^{\alpha\alpha}(0)-c_{-k}^{\alpha\alpha}(0)\right]\bra{k,\alpha}
 J_H\ket{k,\alpha},
 \label{eq:intraJ}
\end{equation}
where $TJ_HT^{-1}=-J_H$ has been used to combine terms with opposite momenta together.    This is zero due to the balance of the $k$ and $-k$ populations, as follows from the time-reversal invariance, Eq.~\eqref{eq:ckTRS}. 

Turning our attention to the interband terms, $\alpha\ne\beta$, we obtain
\begin{equation}
 \braket{J_H}_{\mathrm{inter}}(t)=-i \sum_{k, \alpha\ne\beta}c_{k}^{\alpha\beta}(0)\sin[\omega_{k,\alpha\beta} t]\,\bra{k,\beta}
 J_H\ket{k,\alpha},
 \label{eq:interJ}
\end{equation}
where we used Eq.~\eqref{eq:ckTRS} and $\bra{k,\beta}J_H\ket{k,\alpha}=-\bra{-k,\alpha}J_H\ket{-k,\beta}$, as follows from time-reversal symmetry.  These terms are oscillating in nature.  Hence, for $TAT^{-1}=\pm A$, currents are allowed, but they are either oscillating in time or decaying, due to destructive interference between different frequencies.

To elucidate the role of time-reversal symmetry in preventing DC currents, consider the transition probability for the system in state $\ket{\psi}$ to evolve unitarily for time $t$, be measured, evolve for time $t'$ and become $\ket{\phi}$:
\begin{equation} 
T^{\psi\rightarrow\phi}_{t,t'}=\sum_a|\braket{\phi|e^{-iHt'}P_ae^{-iHt}|\psi}|^2.
\end{equation}
If $THT^{-1}=H$ and $TAT^{-1}=\pm A$ this is equal to the probability of the time-reversed process,
\begin{equation} T^{\tilde{\phi}\rightarrow\tilde{\psi}}_{t',t}=\sum_a|\braket{\tilde{\psi}|e^{-iHt}P_ae^{-iHt'}|\tilde{\phi}}|^2,
\end{equation}
where $\ket{\tilde{\psi}}$ and $\ket{\tilde{\phi}}$ are time-reversed partners of states $\ket{\psi}$ and $\ket{\phi}$.  Hermiticity of the projectors $P_a^\dagger=P_a$ implies that, for $t=t'=0$, $T^{\psi\rightarrow\phi}_{0,0}=T^{\phi\rightarrow\psi}_{0,0}.$  This, combined with $T^{\psi\rightarrow\phi}_{0,0}=T^{\tilde{
\phi}\rightarrow\tilde{\psi}}_{0,0}$ implies that
\begin{equation} T^{\tilde{\phi}\rightarrow\tilde{\psi}}_{0,0}=T^{\phi\rightarrow\psi}_{0,0}.
\label{eq:24}
\end{equation}
As a consequence, if there is a balance of populations between two time-reversed states before a measurement, the balance will remain after the measurement.  This implies the absence of DC currents, see Eq.~\eqref{eq:intraJ}.   If $TAT^{-1}\ne \pm A$, then in general the rates in Eq.~\eqref{eq:24} {\em would differ}.

In sum, measurements of observables that are $T$-eigenoperators cause oscillating or decaying currents when the initial state is time-reversal even.    On the other hand, measurements that are not $T$-eigenoperators can lead to DC currents.

Turning our attention to $\hat{Q}$,  time-reversal symmetry does not prevent a measurement charge transfer since $\hat{Q}$ is even under $T$. On the other hand, $IT$ symmetry does prevent it, see Eq.~\eqref{eq:Q_conditions}.  Below, this will be illustrated in the Rice-Mele model.

As in the previous section, the limitations on the existence of currents are unchanged even if the initial state is not thermal, as long as it is time-reversal symmetric and translationally invariant.  In Appendix~\ref{app:tEven} we show that in the special case when $A$ is time-even and non-degenerate, the conclusions remain valid regardless of the initial state.

\subsection{Summary and numerical illustration}\label{sec:biased}

Table~\ref{tab:symmetry} summarizes the conclusions of the symmetry analysis for the current after a single measurement.  Starting from an equilibrium state, if the Hamiltonian is inversion symmetric and the observable $A$ is an inversion eigenoperator, then no currents or measurement charge displacements develop.  Therefore, currents require breaking of inversion symmetry in either $H$ or $A$.  When one of these occurs, then the time-reversal properties of $A$ determine whether there are DC currents or not, and also whether the Hamiltonian current immediately following a measurement vanishes or not.

\begin{table}[ht]
\begin{center}
\begin{tabular}{c||c|c|c|c|c}
  $IHI^{-1}=H$ & \cmark & \cmark & \cmark & \xmark & \xmark \\
 $IAI^{-1}=\pm A$ & \cmark & \xmark & \xmark & \cmark & \cmark \\ \hline
 $TAT^{-1}=\pm A$ & any & \cmark & \xmark & \cmark & \xmark \\   
     \hline \hline
$\braket{J_H}(0^+)$ & $0$ & $0$ & \notzerop & $0$ & \notzerop \\
$\braket{J_H}(t)$& $0$ & \notzerop & \notzerop & \notzerop & \notzerop \\
$\braket{J_H}_{\mathrm{DC}}$& $0$ &$0$ & \notzerop & $0$ & \notzerop \\
$\braket{\hat{Q}}$ & $0$ & \notzerop & \notzerop$^*$ & \notzerop$^\dagger$ & \notzerop \\
\end{tabular}
\end{center}
 \caption{Constraints on currents arising from spatial inversion and time reversal symmetries, following a measurement starting from the equilibrium state.  The symbols \cmark\, and \xmark\, indicate whether the conditions on the leftmost column are satisfied. A value of $0$ indicates that a given quantity vanishes by symmetry, while \notzerop\, means it is symmetry-allowed. $\braket{J_H}(0^+)$ is the Hamiltonian current immediately after a measurement. The measurement charge displacement $\braket{\hat{Q}}$ is permitted by symmetry if inversion symmetry is disrupted, either by $H$ or $A$. However, two exceptions are observed: ($*$) an observable that is not an eigenoperator of $I$ or $T$, but has $IT$-symmetry will lead to $\langle \hat{Q} \rangle=0$.  In addition, $\langle \hat{Q} \rangle=0$ always for measurements of position ($\dagger$).}
 \label{tab:symmetry}
\end{table}

\begin{figure}
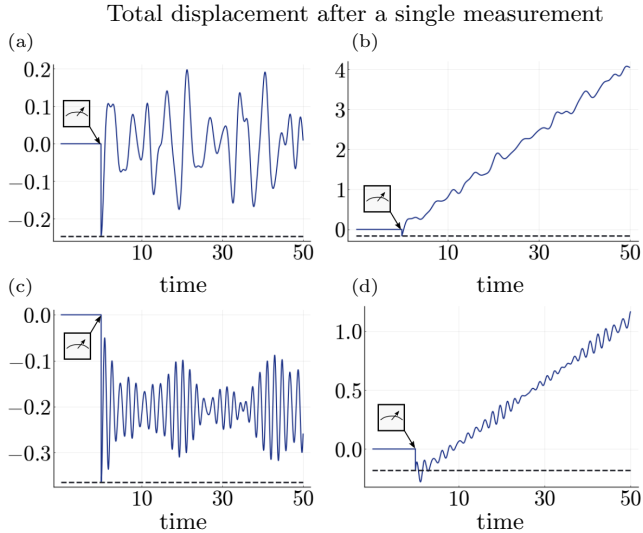

    \centering
    \begin{overpic}
    [width=.49\textwidth]{images/current_dynamics.pdf} 
    \put(17, 76){Total displacement after a single measurement} 
    \put(2, 72){\scriptsize{(a)}} 
    \put(25, 35){time}
    \put(54, 72){\scriptsize{(b)}}
    \put(73, 35){time}
    \put(2, 35){\scriptsize{(c)}}
    \put(25, -1){time}
    \put(54, 35){\scriptsize{(d)}}
    \put(73, -1){time}
\end{overpic} 
\caption{Total displacement as a function of time. The measured observables and staggered potential for the graphs are: (a) $\munit{y}=0,\ \munit{x}=\munit{z}=\frac{1}{\sqrt{2}};\ V=0$; (b) $\munit{x}=\munit{y}=\munit{z}=\frac{1}{\sqrt{3}}, \ V=0$; (c) $\munit{x}=1, \ V=3$; and (d) $\munit{x}=0,\ \munit{y}=-\munit{z}=\frac{1}{\sqrt{2}}, \ V=3 $. The dashed line shows the charge displacement at the instant of measurement. In panels (a) and (b) the Hamiltonian has spatial inversion, but the measured observable breaks it. The oscillating part of the current is always present, while there is also a DC component in (b) when time reversal is broken by the measured observable. In panels (c) and (d) the Hamiltonian doesn't have spatial inversion symmetry, but observables do. If a time-reversal symmetric observable is measured (as in (c)), there is no DC component, while if it is broken as in (d) it is present.}
\label{fig:currentDynamics}
\end{figure}

We illustrate the role of symmetries in constraining the currents using the Rice-Mele model.  Figure~\ref{fig:currentDynamics} shows the total displacement,
\begin{equation}
 \Delta x (t)=\langle \hat{Q}\rangle+\int_0^t dt' \langle J_{\rm{H}}\rangle (t')
\end{equation}
as a function of time for a measurement starting from equilibrium.  Four different cases are considered.  In panels (a) and (b), the Hamiltonian is inversion invariant, $IHI^{-1}=H$, while $IAI^{-1}\neq \pm A$. Conversely, in panels (c) and (d), $IHI^{-1}\neq H$ while $IAI^{-1}= \pm A$. In panels (a) and (c), $TAT^{-1}=\pm A$, whereas in panels (b) and (d), $TAT^{-1}\neq \pm A$. It is important to note that in all scenarios, a measurement displacement occurs ($\langle \hat{Q} \rangle\neq 0$), which is succeeded by oscillatory currents. Furthermore, the disruption of inversion symmetry by either $H$ or $A$ results in similar behavior. However, a qualitative distinction exists between measurements that are not eigenoperators of $T$ and those that are: only measurements that are not $T$-eigenoperators permit DC currents.

\begin{figure}
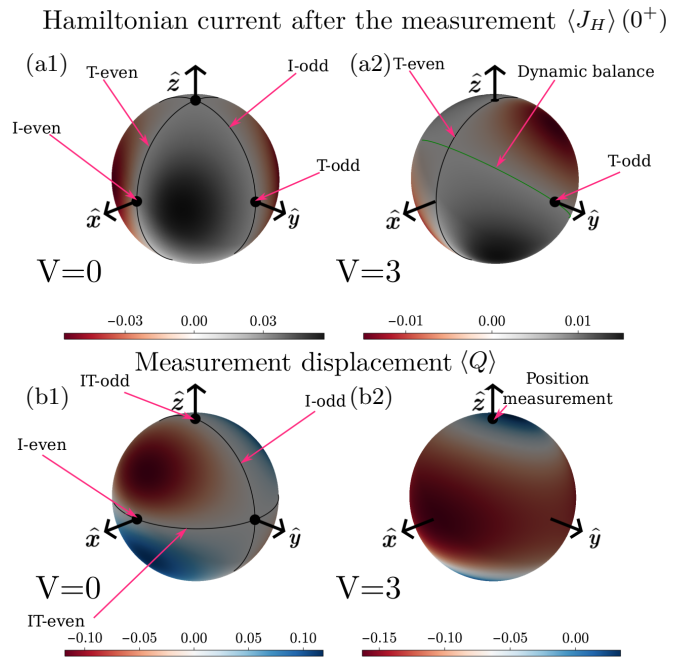

    \centering\begin{overpic}[width=.48\textwidth]{images/four_spheres_current.pdf}  
    \put(2, 90){(a1)}
    \put(50.5, 90){(a2)}
    \put(83, 96.5){$\braket{J_H}(0^+)$}
    \put(2, 40){(b1)}
    \put(50.5, 40){(b2)}
    \put(67, 45.5){$\braket{Q}$}
    \end{overpic}
    \caption{ Transferred charge for every bond observable represented on the Bloch sphere. The parameters of all plots are $t_1=1$, $t_2=0.5$, while the spatial inversion is controlled by a staggered potential $V$. For Hamiltonian current immediately after the measurement, the measurement of $T$ and $I$ eigenoperators leads to the vanishing of the current (shown in (a1)) if Hamiltonian has spatial inversion. In (a2), the non-zero staggered potential breaks spatial inversion.  Then, only time-reversal symmetry gives a simple condition for zero current. The zero-current line shown in green is a result of dynamic balance, rather than symmetry. In (b1) and (b2) we show similar results for the measurement displacement. If the Hamiltonian has a center of symmetry as in (b1), the symmetries to consider are spatial inversion and $IT$-symmetry. If the Hamiltonian breaks this symmetry as in (b2), the only point at which displacement is guaranteed to vanish corresponds to the measurement of position, for which the displacement operator is identically zero.} 
    \label{fig:four_spheres}
\end{figure}

To see the dependence of current on the parameters, Fig.~\ref{fig:four_spheres} shows the currents for a measurement starting from equilibrium, as a function of $\mathbf{\hat{m}}$ on the Bloch sphere. 
Panel (a1) displays the Hamiltonian current {\em immediately following the measurement}, $\langle J_{\mathrm{H}}\rangle (0^+)$, while panel (b1) shows the measurement displacement, $\langle \hat{Q} \rangle$, both in the case where no staggered potential is present, $V=0$.  Then, the Hamiltonian has time-reversal and inversion symmetries, as does the equilibrium state.  As a consequence, currents are only expected for $IAI^{-1}\ne \pm A$.   Time reversal $T$ corresponds to complex conjugation in the position basis, thus changing the sign of $\sigma_{\mathrm{y}}$ while leaving the other two matrices unchanged since they only have real elements.    On the other hand, inversion $I$  exchanges the left and right sites within a cell and is hence represented by $\sigma_{\mathrm{x}}$.  $IT$ symmetry is a combination of these two symmetries and therefore can be represented as $\sigma_{\mathrm{x}}K$, where $K$ is a complex conjugation operator.  The current and measurement displacement vanish along the planes and axes of these symmetries.

Panels (a2) and (b2) show $\langle J_{\mathrm{H}}\rangle (0^+)$ and $\langle \hat{Q}\rangle$ for $V\neq 0$, when the equilibrium state is not inversion invariant.  The locus of points that are $T$-even or $T$-odd still have vanishing Hamiltonian currents $\langle J_{\mathrm{H}}\rangle (0^+)$ as shown in (a2). However, inversion is no longer a useful guide to determine the locations of zero current.  In this case, the current vanishes when the inversion breaking of the observable and that of the staggered potential counteract each other. For the measurement displacement in panel (b2), when inversion is broken,  $IT$-symmetry is also broken as default. Then, the only point at which the measurement displacement is guaranteed to vanish is the position measurement $m_z=1$ for which the displacement is identically zero. 

\section{Steady states of thermally isolated system}\label{sec:InfiniteT}

We now turn our attention to systems that are measured repeatedly.  We consider multiple measurements of the same observable $A$ that occur at a series of arbitrary times.  If this is continued for a sufficiently long time, then under fairly generic conditions, the system reaches an infinite-temperature steady state~\cite{neumann1996, lindblad1972, lindblad1973, skinner2019, yi2011}.  We show that currents vanish in this state irrespective of symmetry considerations.

To see that the steady state of a measured quantum system is an infinite-temperature state, we recall that a projective measurement when averaged over all possible outcomes causes the von Neumann entropy of a system $S(\rho)=-\mathrm{tr}[\rho\log\rho]$ to increase (or stay the same)~\cite{lindblad1972, ruelle1999}.
The infinite-temperature density matrix, being proportional to the identity matrix, is a steady state of both unitary evolution ($U\mathbb{1}U^{\dagger}=\mathbb{1}$) and of projective measurements ($\sum_aP_a\mathbb{1}P_a=\mathbb{1}$). Therefore, to the extent that the dynamics is ergodic, the steady state under repeated measurements has infinite temperature.  This fate can be avoided by two different mechanisms for non-ergodicity.  First, if the observable $A$ commutes with the Hamiltonian, states with different eigenvalues of $A$ are prevented from mixing during the evolution.  Second, when measurements are performed infinitely fast, they effectively freeze the unitary evolution between measurements leading to the quantum Zeno effect~\cite{facchi2008}.

The tendency of measurements to lead the system toward infinite temperature can be further clarified by looking at the transition rates that are induced by it. For two energy eigenstates $\ket{E_1}$ and $\ket{E_2}$, the transition probabilities caused by the measurement of an observable $A$ are identical,
\begin{equation}
    \sum_a|\braket{E_1|P_a|E_2}|^2=\sum_a |\braket{E_2|P_a|E_1}|^2.    
\end{equation}
This implies that the steady state has equal occupation of energy eigenstates, independent of their energy.

\subsection{Absence of currents in infinite temperature}

Given that the steady state is generically an infinite-temperature state, we now show that currents vanish in this state. There are well-known statistical mechanics systems, such as the Asymmetric Simple Exclusion Process (ASEP)~\cite{ASEP}, which support steady-state currents even with a uniform probability distribution. In this case, the currents arise due to asymmetry in the hopping amplitudes, and one may expect that inversion-breaking measurements could also generate currents through a similar mechanism.  We now show that measurements cannot lead to such currents.

We first consider the Hamiltonian current.  Writing the infinite-temperature density matrix as $\rho_{\infty}=\frac{1}{\mathcal{N}}\mathbb{1}$ (here, $\mathcal{N}$ is the dimensionality of the Hilbert space), the expectation value of the Hamiltonian current is

\begin{equation}\label{eq:infT_ham}
    \braket{J_H}_{\infty}=\frac{1}{\mathcal{N}}\mathrm{tr}\left(-i[H,\hat{x}]\right)=0,
\end{equation}
as follows from the cyclic property of the trace. 

We turn our attention to the measurement charge displacement.  In the infinite-temperature state, $\rho=\mathbb{1}/\mathcal{N}$, 
\begin{equation}
\braket{\hat{Q}}_{\infty}=\frac{1}{\mathcal{N}}\mathrm{tr}\left(\hat{x}\sum_{a}P_{a}P_a\right)-\frac{1}{\mathcal{N}}\mathrm{tr}\left(\hat{x}\right).
\end{equation}
where we used the cyclic property of the trace.  This is seen to vanish identically, since $\sum_a P_a^2=\sum_a P_a=\mathbb{1}$. Hence, projective measurement cannot result in charge transfer at an infinite temperature.
\footnote{It is straightforward to check that this argument can be generalized to any Kraus operator
$\rho \mapsto \rho' = \sum_{a} K_a \rho K_a^{\dagger}$ that is normal, $[K_a,K_a^\dagger]=0$.
}

A natural way to avoid the infinite-temperature state is to couple the system to a heat bath.  This allows the system to reach a non-trivial steady state that emerges as a consequence of a balance between the bath that cools the system and the measurement apparatus that heats it up.  We consider this in the next section.

\section{Coupling to a thermal bath}

\label{sec:thermalBath}

We model the dissipative dynamics through the Markovian quantum master equation (Lindblad equation) where the dynamics of an open quantum system is generated by the Lindbladian $\mathcal{L}_0$,
\begin{equation}\label{eq:lindbladian0}
    \mathcal{L}_\mathrm{HD}[\rho]=-i[H,\rho]+\mathcal{D}[\rho]\,,
\end{equation}
where $\mathcal{D}$ is the dissipator which accounts for the coupling to a thermal bath,
\begin{align}\label{eq:dissipator}
    \mathcal{D}[\rho]=\sum_{\alpha}\left(L_{\alpha}\rho L_\alpha^{\dagger}-\frac{1}{2}\{L_\alpha^{\dagger}L_\alpha, \rho\}\right)\,.   
\end{align} 
We choose jump operators $L_\alpha$ so that, in the absence of measurements, they thermalize the system at a temperature $T$. This means that the steady state of this Lindbladian is a Gibbs state $\rho_{\mathrm{eq}}=e^{-\frac{H}{T}}/Z$. 

We now imagine that the system is being measured. We start by considering measurements that occur at random uncorrelated times (a Poisson process) with a measurement rate $1/\tau$~\cite{dum1992, jacobs2006, jacobs2014}.  Later, in Sec.~\ref{sec:Periodic}, we will consider time-periodic measurements. 
In this case, the measurement process can be described by introducing an additional term in the Lindbladian (see Appendix~\ref{app:measurementLindbladian}),
\begin{equation}\label{eq:full_lindbladian}
    \mathcal{L}=\mathcal{L}_\mathrm{HD}+\frac{1}{\tau}\mathcal{L}_{\mathrm{m}},
\end{equation}
where $\mathcal{L}_{\mathrm{m}}$ acts on the density matrix as 
\begin{equation}
    \mathcal{L}_{\mathrm{m}}[\rho]=\sum_a\left(P_a\rho P_a\right)-\rho.
\end{equation}
 We observe that the Kraus map $\mathcal{P}_A[\rho]=\sum_a P_a\rho P_a$, corresponding to measurements is a super-projector. We can define the complementary projector $\mathcal{Q}_A=1-\mathcal{P}_A$, so that $\mathcal{P}_A\mathcal{Q}_A=0$. The measurement term in the Lindbladian (Eq.~\ref{eq:full_lindbladian}) is $\frac{1}{\tau}\mathcal{L}_{\mathrm{m}}\rho = -\frac{1}{\tau}\mathcal{Q}_A\rho$ which causes any elements that are off-diagonal in the measurement basis to decay with a characteristic decay time $\tau$.

It is interesting to note that the same Lindbladian can be obtained for a system under continuous weak measurement of an observable $A$, see Ref.~\onlinecite{jacobs2006}. There, the measurement strength takes on the role of the measurement rate $1/\tau$ in our analysis.  Therefore, our discussion of currents in the Poisson measurement scheme is also relevant to such systems.

The steady state can be determined from the condition $\mathcal{L}[\rho_{\mathrm{st}}]=0$, or 
\begin{equation}\label{eq:random_lindbladian}
    \left(\mathcal{L}_{\mathrm{m}}+\tau\mathcal{L}_\mathrm{HD}\right)[\rho_{\mathrm{st}}]=0.
\end{equation}
It is interesting to see how the Lindblad equation predicts an infinite temperature state in the absence of dissipation.  Then, Eq.~\eqref{eq:random_lindbladian} becomes,
\begin{equation}\label{eq:ssNoDissipation}
-\calQ_A\rho_\mathrm{st}-i\tau\left[H,\rho_\mathrm{st}\right]=0\,.
\end{equation}
Multiplying this by $\rho_\mathrm{st}$ and computing the trace gives $0=-i\tau\ \mathrm{tr}\left[\rho_\mathrm{st}[H,\rho_\mathrm{st}]\right]=\mathrm{tr} \left[\rho_\mathrm{st}\calQ_A\rho_\mathrm{st}\right]=\mathrm{tr} \left[(\calQ_A\rho_\mathrm{st})^2\right]$, where we used $\mathrm{tr}\left[(\calP_A\rho_\mathrm{st})(\calQ_A\rho_\mathrm{st})\right]=0$.  This implies $\calQ_A\rho_\mathrm{st}=0$, {\em i.e.} $\rho_\mathrm{st}$ commutes with $A$. If $\calQ_A\rho_\mathrm{st}=0$, then Eq.~\eqref{eq:ssNoDissipation} implies that $\rho_\mathrm{st}$ commutes with $H$ as well.  For generic $A$ and $H$, $\rho_\mathrm{st}$ must therefore be the infinite temperature state.
\footnote{This relies on the result that if $A$ and $H$ are a generic pair of Hermitian matrices, there is no Hermitian matrix besides a scalar matrix that commutes with both of them.
Let $M$ commute with both, and let $\lambda$ one of its eigenvalues.  Let $V$ be the span of all eigenvectors of $M$ with this eigenvalue, which will be a proper subspace if $M$ is not a scalar matrix.  $A$ maps $V$ into itself, since it commutes with $M$.
This implies that $A$ can be diagonalized within the space $V$, so $V$ is the span of a set of $A$'s eigenvectors.
This is also true for $H$.  So there is a combination of eigenvectors of $A$ and a combination of eigenvectors of $H$ that span the same space, which is not a generic situation.}

We now turn to define and study the steady-state currents. 

\subsection{Currents in Measured Open Quantum Systems}

In an open quantum system undergoing Lindblad dynamics, the Hamiltonian current does not satisfy the continuity equation.  The presence of measurements and dissipation in the master equation forces us to consider new types of currents: measurement currents and dissipative currents.  Only the total current, which is the sum of all three, is conserved. As we will show below, the measurement current is directly related to the measurement charge displacement $\hat{Q}$.

We use the generalized definition of current between sites $x$ and $y$ for a quantum master equation \cite{hovhannisyan2019},
\begin{equation}\label{eq:current_definition}
    j^{x\rightarrow y} = \frac{1}{2}\left\{P_x, \frac{dP_y}{dt}\right\}-\frac{1}{2}\left\{P_y, \frac{d P_x}{dt}\right\},
\end{equation}
where $P_x$ and $P_y$ are projectors onto $x$ and $y$ and $\{,\}$ is the anticommutator. The current is defined so that it explicitly satisfies the continuity equation.  The current, integrated over all space, is then
\begin{equation}\label{eq:integrated_current}
    J=\sum_{x<y}(y-x)j^{x\to y}
\end{equation}
where the factor of $(y-x)$ takes into account the 
distance traveled by a particle in the transition from $x$ to $y$\footnote{For periodic boundary conditions, $(y-x)$ is the shortest displacement on the ring.}. Equation (\ref{eq:current_definition}) is a Heisenberg picture expression, written in terms of the time derivatives of the projectors $P_x$.  A more explicit expression for the current can be obtained using \cite{hovhannisyan2019}
\begin{equation}
    \frac{d}{dt} P_x=\mathcal{L}^\dagger\left[P_x\right]
\end{equation}
where $\mathcal{L}^\dagger$ is the adjoint Lindbladian,
\begin{align}\label{eq:adjointLind}
    \mathcal{L}^\dagger\left[\Theta\right]=&i[H,\Theta]+\frac{1}{\tau}\left(\sum_a\left(P_a\Theta P_a\right)-\Theta\right)\nonumber\\&+\sum_\alpha \left(L_\alpha^\dagger\Theta L_\alpha-\frac{1}{2}\{L_\alpha^\dagger L_\alpha,\Theta\}\right)\,.
\end{align}
Replacing $\frac{dP_x}{dt}$ by $\mathcal{L}^\dagger[P_x]$ (and similarly for $P_y$) in Eq.~(\ref{eq:current_definition}) gives the expressions for the Hamiltonian, measurement, and dissipative currents, depending on which term of the adjoint Lindbladian is used.

The Hamiltonian part of the evolution leads to the \emph{Hamiltonian current}, discussed in Sec.~\ref{sec:definitions}. The space-integrated current takes the form
\begin{equation}
    J_H = - i\sum_{x, y} P_y HP_x\left(y-x\right)=-i[\hat{x}, H]\,.
\label{eq:TotalJH}
\end{equation}

The measurement part of the Lindbladian $\mathcal{L}_{\mathrm{m}}$ gives the \emph{measurement current},
\begin{equation}\label{eq:measurement_current_definition}
    j_{\mathrm{meas}}^{x\rightarrow y}=\frac{1}{2\tau}\sum_a\left(\{P_x, P_a P_y P_a\}-\left(x\leftrightarrow y\right)\right).
\end{equation}
Integrating over space gives
\begin{equation}
    J_\mathrm{meas} = \frac{1}{\tau} \left(\sum_a P_a\hat{x}P_a-\hat{x}\right)=\frac{\hat{Q}}{\tau}\,,
\label{eq:TotalJm}
\end{equation}
where $\hat{Q}$ is defined in Eq.~(\ref{eq:Qx}). This relation describes the fact that charge displacements occur on average at a rate $1/\tau$.

Finally, the last term in Eq.~\eqref{eq:adjointLind} is a source of a \emph{dissipative current} that comes from thermal bath jump operators \cite{hovhannisyan2019}.  This gives
\begin{equation}
\begin{split}
    J_{\mathrm{dis}}&=\frac{1}{2}\sum_{n,m}j_{\mathrm{dis}}^{n\rightarrow m}\left(m-n\right) \\
    &=\sum_{\alpha}\left[L_{\alpha}^{\dagger}\hat{x}L_{\alpha}-\frac{1}{2}\{\hat{x}, L_{\alpha}^{\dagger}L_{\alpha}\}\right].
\end{split}
\end{equation}
The second equality relies on the locality of $j_{\mathrm{dis}}^{n\rightarrow m}$.  For the Hamiltonian and measurement currents, locality is guaranteed by the short-range interactions and local measurements.  By contrast, to ensure equilibration it is often convenient to work with a bath that spans the whole system, rendering the definition of dissipative currents problematic.  For this reason, we focus on a sufficiently weak dissipation rate, compared to the Hamiltonian scale and to the measurement rate $1/\tau$.  Then, the dissipative current can be neglected.  We comment that the use of a Lindblad master equation to describe the coupling to a thermal bath is only justified in this limit~\cite{spohn1978}.

\subsection{Measurement-induced steady-state currents in the Rice-Mele model}\label{sec:example}

We will demonstrate the existence and properties of steady-state currents on the Rice-Mele model defined in the introduction.  As will become clear, much of the intuition comes from the single-measurement results.  We model the thermal bath by choosing jump operators that connect energy eigenstates, 
\begin{equation}
L_\alpha=L_{k,\mu;q,\nu}=\sqrt{\gamma_{k,\mu; q,\nu}}c_{k,\mu}^{\dagger}c_{q, \nu} \, ,
\end{equation} with the decay rates chosen to satisfy detailed balance at temperature $T$,
\begin{equation}
    \gamma_{k,\mu; q,\nu}=\begin{cases}
    \frac{\gamma_0}{\mathcal{N}} \hspace{3mm} &\mathrm{for}\ \ E_{k,\mu} < E_{q,\nu};\\
    \frac{\gamma_0}{\mathcal{N}}e^{-\frac{E_{k,\mu}- E_{q,\nu}}{T}} &\mathrm{for}\ \  E_{k,\mu} \geq E_{q,\nu};
    \end{cases}
\end{equation}
where $k$ and $q$ are wavevectors, $\mu$ and $\nu$ band indices, and $\gamma_0$ sets the overall timescale of relaxation. The factor $1/\mathcal{N}$ is added since we allow transitions between all pairs of energy levels. This ensures that the relaxation rate of the system remains finite as $\mathcal{N}\to\infty$.  The dissipation is inversion symmetric if the Hamiltonian is, so we do not have to consider its symmetry properties independently.

The steady state of the Lindbladian -- consisting of the Hamiltonian dynamics together with dissipation and measurements -- is found numerically.  Throughout we assume a weak dissipation rate compared to the Hamiltonian scale, $\gamma_0\ll ||H||$.

\begin{figure}
     \centering
\subfloat{
\begin{overpic}
[width=.42\textwidth]{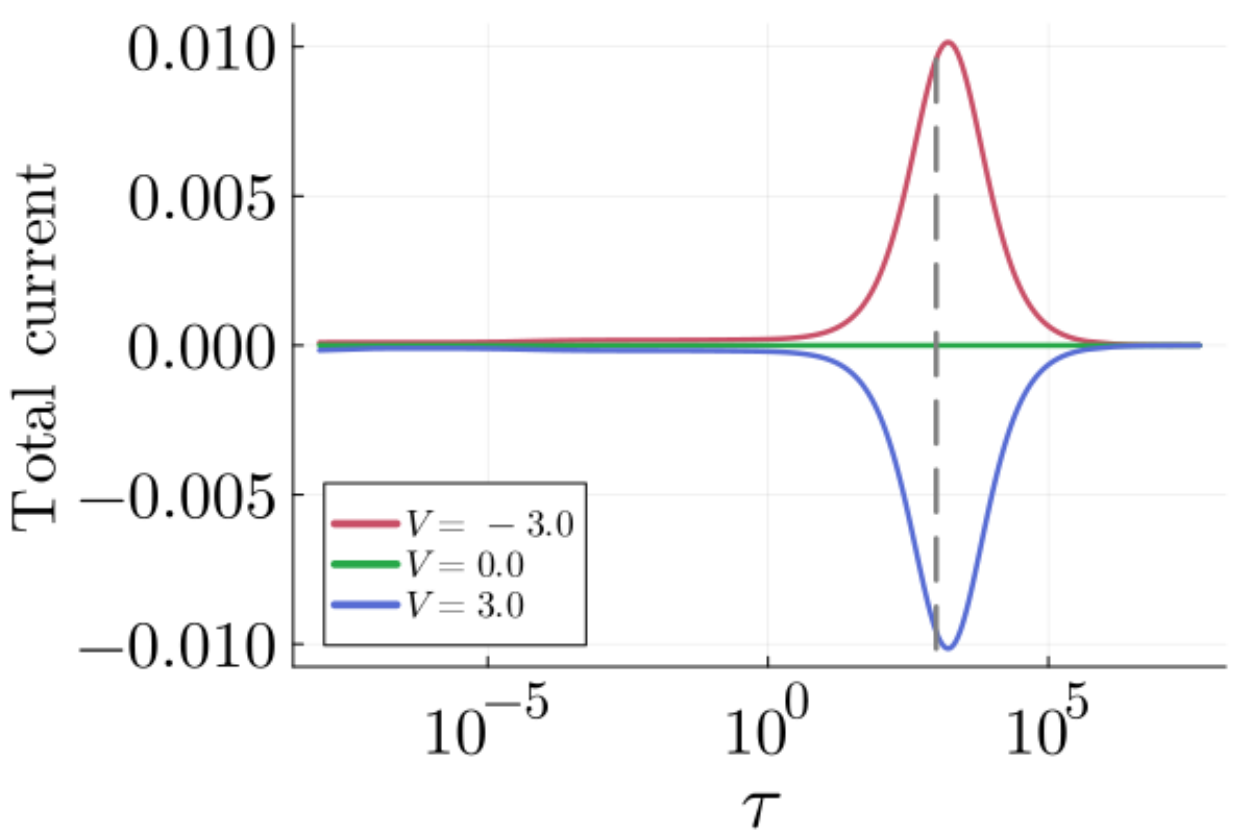}
\put(3, 65){(a)}
\put(70, 10){$1/\gamma_0$}
\put(24, 43){\includegraphics[scale=0.13]{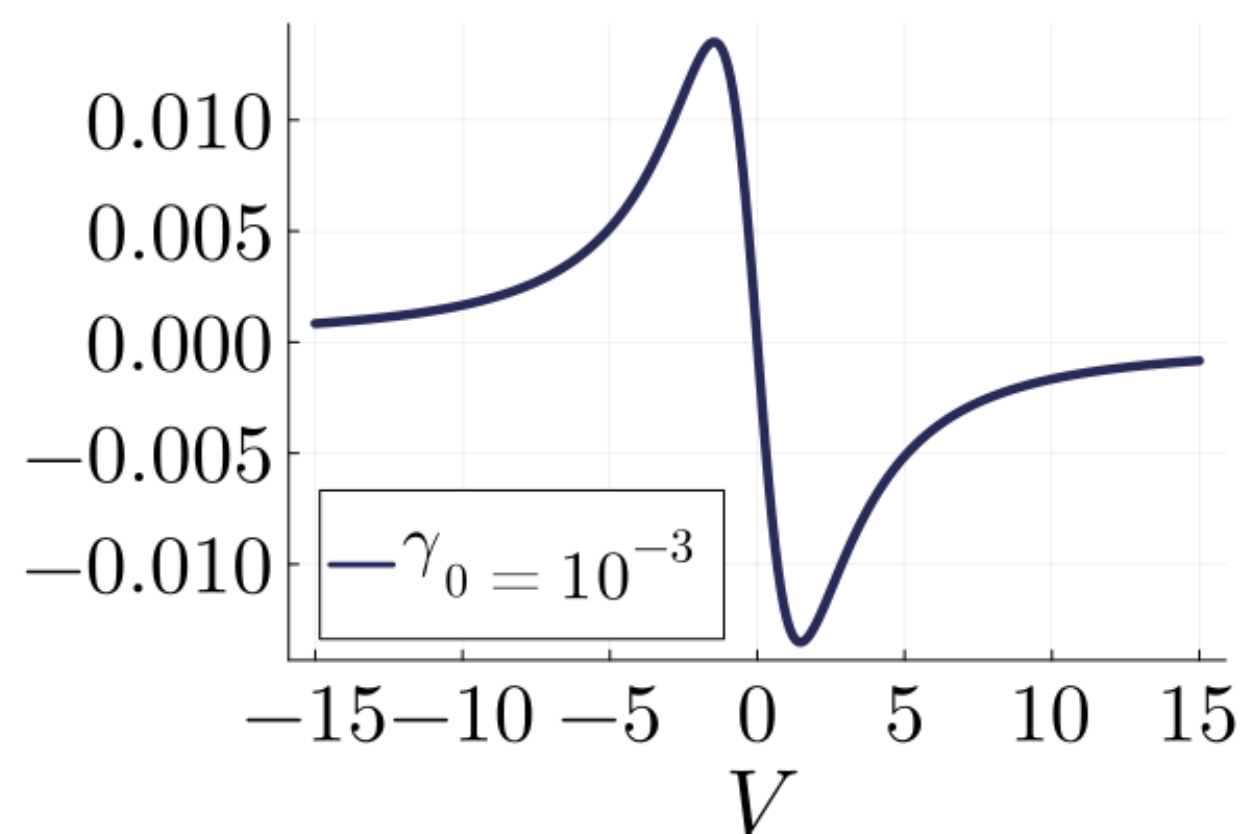}}
\end{overpic} 
\label{fig:currenta}}\\
\subfloat{\begin{overpic}[width=.42\textwidth]{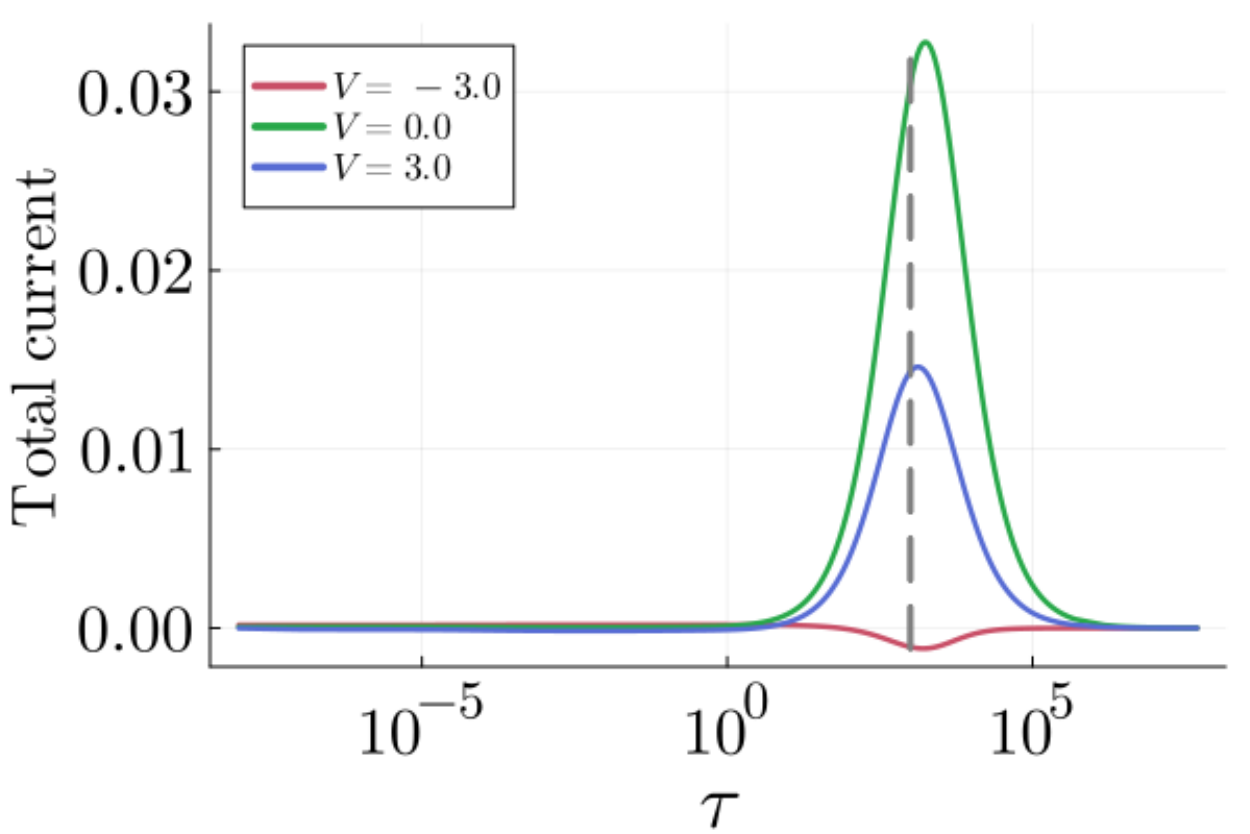}
\put(3, 65)
{(b)}
\put(70, 10){$1/\gamma_0$}
\put(22, 21){\includegraphics[scale=0.13]{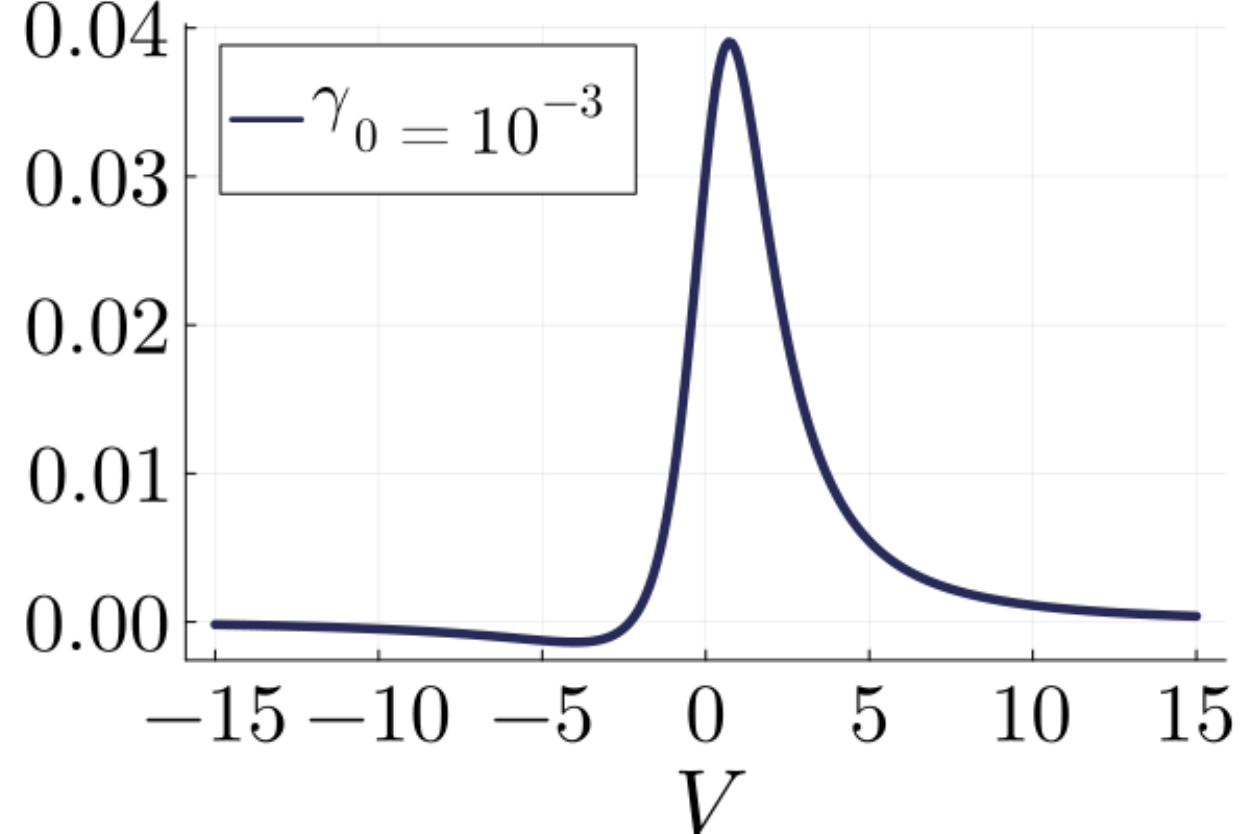}}
\end{overpic}
\label{fig:currentb}}\\
\subfloat{\begin{overpic}[width=.42\textwidth]{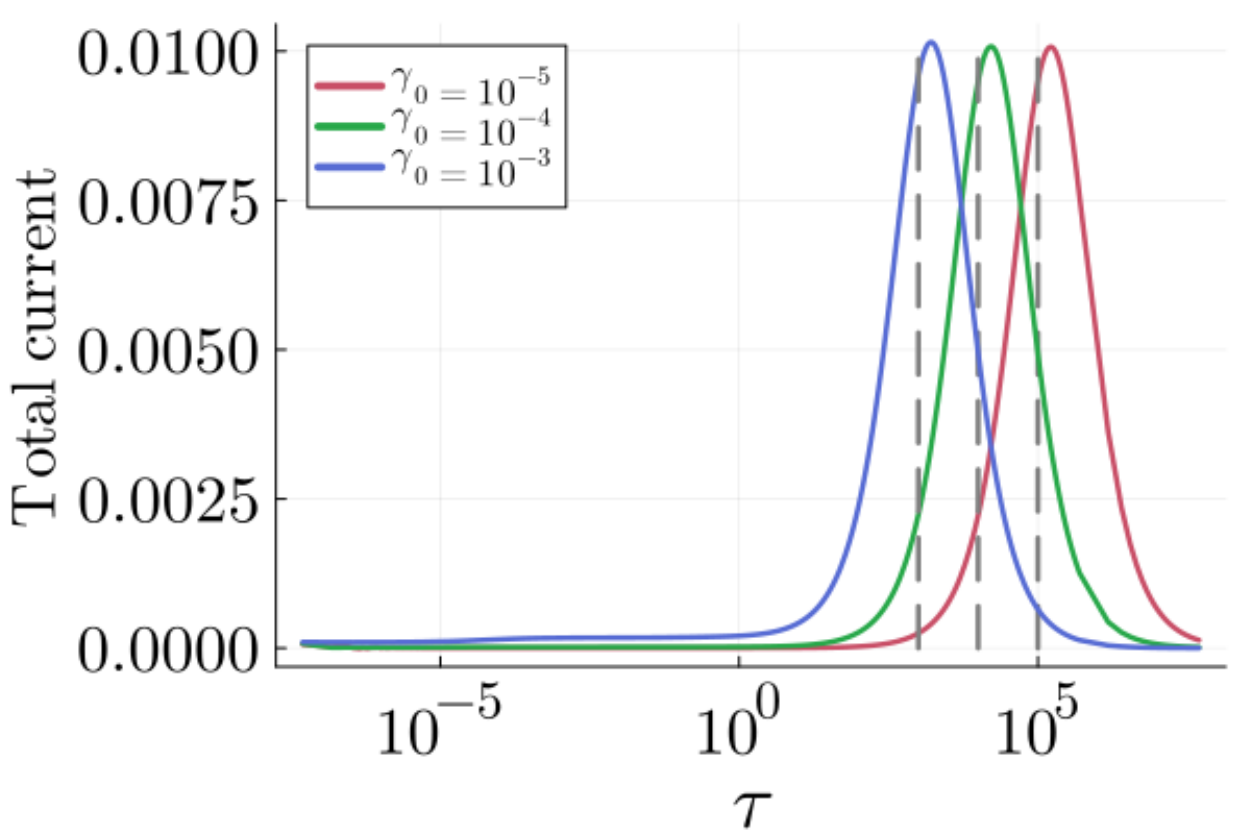}
\put(3, 67){(c)}
\end{overpic}         \label{fig:currentc}
}
\caption{Steady-state currents as a function of average measurement time.  In panel (a) the measured observable is inversion-odd and it is given by $m_y=m_z=1/\sqrt{2}$, and $m_x=0$. Note that the $\tau$-axis is logarithmic in all plots. The total current dependence is shown for three values of the staggered potential as the staggered potential $V$ is varied. The relaxation scale, $\tau\approx1/\gamma_0$, is marked by dashed lines, where $\gamma_0=10^{-3}$. The current changes direction when the staggered potential is tilted in the opposite direction ($V\rightarrow -V$). This is clearly not the case when the measured observable is not an inversion eigenoperator, as seen in panel (b) where the measured observable is given by $m_x=m_y=-m_z=1/\sqrt{3}$. In insets we plot the dependence of total current on staggered potential for $\tau=1/\gamma_0$. In panel (c) we show how the location of the peaks matches the relaxation time scale $1/\gamma_0$ where the total current is plotted for different values of $\gamma_0$ and $V=-3.0$. The plot is given for the same inversion-odd observable as the one in panel (a).  The parameters of all the plots are $t_1=1.0$, $t_2=0.5$, $T=0.1$, and $N=3$.}
\label{fig:current_tau}
\end{figure}

Figure \ref{fig:current_tau}(a) shows the total (Hamiltonian plus measurement) current for an inversion-odd observable, as a function of $\tau$. The current is obtained from the expectation value of the corresponding operators with respect to the steady-state density matrix.  The presence of steady-state currents is a demonstration of \emph{the quantum-measurement ratchet effect}, where the combination of an inversion-invariant measurement and an inversion-breaking Hamiltonian give rise to currents. The ratchet effect occurs only for $V\neq0$ since a non-zero staggered potential (ratchet potential) is necessary to break the inversion symmetry. In addition, changing the sign of the potential reverses the direction of the current.

By contrast, Fig.~\ref{fig:current_tau}(b) shows the total current for an observable that is not an inversion eigenoperator, $m_x=m_y=-m_z=1/\sqrt{3}$. Here, a current is created even in the absence of a staggered potential, and the current is not odd in $V$.

The currents in Fig.~\ref{fig:current_tau} display a pronounced peak at $\tau\approx 1/\gamma_0$, which is the typical equilibration time of the bath.  To understand the origin of this peak, we begin from the rare measurement limit $\tau \gg 1/\gamma_0$.  There, the current is expected to approach zero, since the system is only rarely perturbed from equilibrium.  Then, the currents are given by the cumulative displacement following a single measurement before returning to equilibrium, divided by the time between measurements $\tau$.  

In this context, it is important to distinguish between measurements that are time-reversal eigenoperators, $TAT^{-1}=\pm A$, and those that are not.  In the latter case, measurements starting from equilibrium give rise to DC currents, $J_{DC}$.  Due to dissipation, such currents decay on a time scale of order $1/\gamma_0$.  Hence, the cumulative displacement per measurement is $\approx J_{DC}/\gamma_0$, and the average current in the rare-measurement limit is $\approx J_{DC}/(\gamma_0\tau)$. As $\tau$ is decreased from the rare measurement limit, significant heating occurs as a result of the measurements, as will be shown below.  This heating suppresses the currents, an effect that becomes important when measurements are closer together than the relaxation time $\tau_\mathrm{r}=1/\gamma_0$.  The net effect is that a peak in the current is obtained at $\tau\approx \tau_\mathrm{r}$, whose height is $\approx J_{DC}/(\gamma_0\tau_\mathrm{r})= J_{DC}$, which is independent of $\gamma_0$, as seen in Fig.~\ref{fig:current_tau}(c).  

By contrast, for $TAT^{-1}=\pm A$, no DC currents are created starting from equilibrium.  This implies that, in the rare measurement limit, the charge displacement between each pair of measurements is of order $1$, and that the average current is of order $1/\tau$.   Then, the current for $\tau=\tau_\mathrm{r}$ is of order $1/\tau_\mathrm{r}=\gamma_0$, which is negligible for weak dissipation.   Furthermore, as $\tau$ is decreased below $\tau_\mathrm{r}$, the heating is found to prevent currents from developing. For this reason, in order to induce sizable steady-state currents in the weak dissipation limit, measurements must not be time-reversal eigenoperators.

Irrespective of the heating, the current is also expected to approach zero in the Zeno limit, $\tau\ll \tau_\mathrm{Z}$, where $\tau_\mathrm{Z}=\gamma_0/||H||^2$ is the Zeno onset scale, derived in Appendix~\ref{app:perturbation}.  Here, $||H||$ is the characteristic energy scale of the system.  
The emergence of the Zeno onset scale can also be understood intuitively. When $A$ is measured twice in close succession, the system is with high probability determined to remain in the same state.  The Zeno effect comes about since the Born rule implies a quadratic time-scaling of the effective evolution -- which is of order $||H||^2\tau^2$~\cite{facchi2008}. The dissipation, on the other hand, evolves the density matrix according to classical probability, which means that it has a usual linear scaling $\gamma_0\tau$. For sufficiently small $\tau$, the dissipation always wins in this competition, and the Hamiltonian evolution is irrelevant. The turnaround point is at $\tau_{\mathrm{Z}}$ which is the mean measurement time at which the two scales become comparable. For larger times the evolution is dominated by the Hamiltonian, which together with the measurements drive the system toward the infinite-temperature state since the dissipation can be neglected. Then the conclusions of Sec.~\ref{sec:InfiniteT} apply.

In the Zeno limit, currents are small because rapid local measurements tend to freeze particle motion, as will be analyzed in detail in the following section.  

\begin{figure}
     \centering
\subfloat{    \begin{overpic}[width=0.48\textwidth]{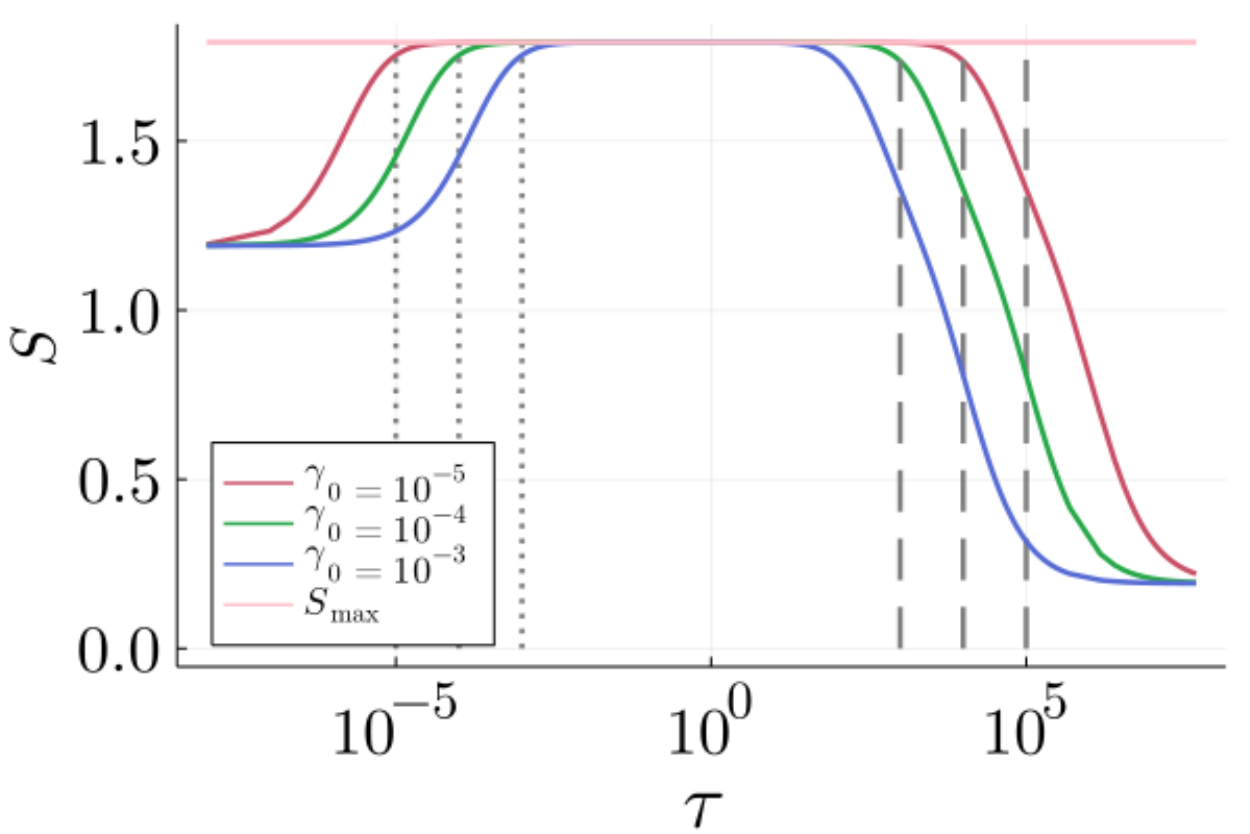}
\put(3, 65){(a)}
\put(38, 10){$\tau_\mathrm{Z}$}
\put(70, 10){$1/\gamma_0$}
\end{overpic}
\label{fig:entropy1}}\\
\subfloat{\begin{overpic}[width=.48\textwidth]{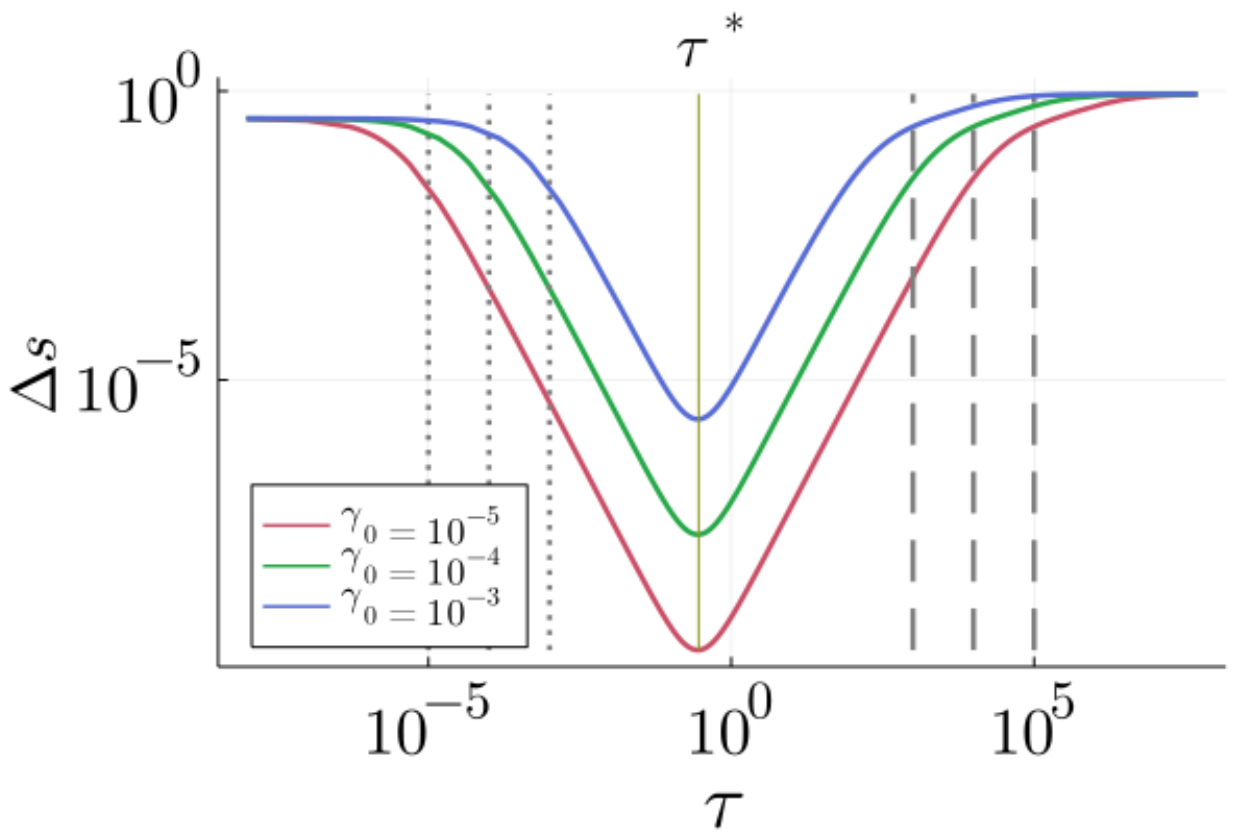}
\put(3, 65){(b)}
\end{overpic} \label{fig:entropy2}}
        \caption{The von Neumann entropy as a function of average measurement time. The measured observable is given by $m_y=m_z=1/\sqrt{2}$, and $m_x=0$. In panel (a) we show the entropy as a function of measurement time $\tau$. The maximal entropy $S_{\mathrm{max}}$ of an infinite-temperature state is almost saturated near the middle of the plot. In panel (b) the relative distance of the entropy from the maximal entropy (Eq. \eqref{eq:relative_entropy}) is plotted on the log-log scale.  The parameters of the plots are the same as in Fig.~\ref{fig:current_tau}. }
    \label{fig:entropy_tau}
\end{figure}

To further understand the suppression of currents below the peak at $\tau\approx 1/\gamma_0$, it is useful to study the von Neumann entropy $S$ of the steady-state density matrix as a function of $\tau$, see Fig.~\ref{fig:entropy_tau}(a).  For rare measurements, $\tau\gg 1/\gamma_0$, the system has time to equilibrate to the bath temperature, which is low in our simulation.  Hence, the entropy becomes small. As $\tau$ is decreased, when the measurement time becomes comparable to the relaxation time $1/\gamma_0$, $S$ rises rapidly.  For $\tau_\mathrm{Z} \ll \tau \ll 1/\gamma_0$, the bath struggles to remove the entropy created by the measurements, and therefore the entropy plateaus close to the infinite temperature value $\log (2N)$.  This is the origin of the current suppression, since the current is guaranteed to vanish in the infinite-temperature state, see Sec.~\ref{sec:InfiniteT}.  Finally, for $\tau\ll \tau_\mathrm{Z}$, the system enters the Zeno limit, in which the Hamiltonian has no effect and the steady state is determined by the interplay between measurements and dissipation. Then, the entropy saturates below the infinite temperature entropy, at a value that is independent of $\gamma_0$.  

Note that as the dissipation is made weaker, the high-entropy plateau in Fig.~\ref{fig:entropy_tau}(a) becomes broader.  This is consistent with the fact that, in the absence of dissipation, the steady state is an infinite temperature state, as discussed in Sec.~\ref{sec:InfiniteT}. At a finite value of $\gamma_0$ the range of the plateau extends from $\tau_Z$ to $1/\gamma_0$, as argued above. The distance to the infinite temperature state can be seen more clearly from the normalized difference between the von Neumann entropy of the system $S$, and the maximum entropy $S_{\mathrm{max}}=\mathrm{log}(2N)$.  The normalized difference,
\begin{equation}\label{eq:relative_entropy}
    \Delta s= \frac{S_{\mathrm{max}}-S}{S_{\mathrm{max}}},
\end{equation}
is plotted as a function of measurement time $\tau$ on a log-log scale in Fig.~\ref{fig:entropy_tau}(b). 
We see that the steady-state entropy reaches a maximum at $\tau^*$, which is of order $1/||H||$ and independent of $\gamma_0$. On either side of $\tau^*$, the maximum entropy displays a power-law behavior over a large window, with $\Delta s\propto \tau^{-2}$ for $\tau_{\mathrm{Z}}\ll \tau\ll\tau^*$ and $\Delta s\propto \tau^{+2}$ for $\tau^*\ll\tau\ll 1/\gamma_0$. An analytic derivation for this behavior is given in Appendix~\ref{app:perturbation}.

\subsection{The Zeno limit and local current loops}
\label{sec:Zeno}

In the Zeno limit measurements are fast relative to the Hamiltonian and dissipative timescales~\footnote{Note that the choice of dissipator introduced in Eq.~\eqref{eq:dissipator} is physically not well-justified in the Zeno limit, when measurements are performed faster than the characteristic timescale of the Hamiltonian.  However, we expect the qualitative behavior not to be affected by the exact model of dissipation.}.  Then the system stays very close to the measurement subspace (the space of density matrices that are block-diagonal in the measurement basis). Therefore, if measurements are local and non-overlapping in space, charge cannot escape the measurement subspace, thus forbidding global currents.   Interestingly, we will show that the Hamiltonian and measurement charge displacements do not individually vanish but their sum does.  Furthermore, despite the absence of charge displacements, we will show that in the Zeno limit steady-state currents can form loops within each measurement subspace, even when the measurement subspace is non-degenerate. These current loops show that motion can occur when a system is continually observed, in a counterpoint to the Zeno effect. While the Zeno effect implies only that transitions between states are very slow, a current of order one is still found.

For simplicity, throughout we assume that the measurement operator $A$ is non-degenerate.  When this is not the case, one may expect interesting dynamics within the degenerate Zeno subspaces~\cite{blumenthal2022, popkov2018, facchi2008,  grunbaum2013, yin2024, dhar2015}, which we do not consider here. 

In the Zeno limit, the density matrix will be nearly diagonal in the basis of measurement outcomes.  The probabilities of the outcomes are determined by balancing transitions between them by the coupling to the bath. To represent this analytically we treat the measurements as the dominant part of the Lindbladian, and the unitary evolution and the coupling to the bath as perturbations.  Then, using the perturbation theory derived in Appendix~\ref{app:perturbation}, we find that to zeroth order in $\tau$,  
\begin{eqnarray}
\rho_{\mathrm{st}}^{(0)}=\sum_a w_a^{(0)}\ket{a}\bra{a}\,,
\label{eq:zeroth_rho}
\end{eqnarray}
where $w_a^{(0)}$ is the occupation of the eigenstate $\ket{a}$.  This is given by the balance equation
\begin{equation}\label{eq:balance_equation}
    \sum_{a'\neq a}w_{a'}^{(0)}T^{a'\rightarrow a}=w_a^{(0)}\sum_{a'\neq a}T^{a\rightarrow a'},
\end{equation}
where $T^{a\rightarrow a'}=\sum_\alpha |\braket{a'|L_\alpha|a}|^2$ is the transition rate from the state $\ket{a}$ to the state $\ket{a'}$ and $L_\alpha$ are the quantum jump operators, see Eq.~\eqref{eq:dissipator}. Note that, even though the dissipation is weak, it determines the steady state in the strict Zeno limit, whereas the Hamiltonian appears only at higher orders in $\tau$.

\subsubsection{Hamiltonian and measurement currents in the Zeno limit}

Since the density matrix in Eq.~\eqref{eq:zeroth_rho} is diagonal in the measurement basis it is unaffected by measurements.  Therefore, the expectation value of the measurement current with respect to the zeroth order density matrix vanishes, $\langle J_\mathrm{meas} \rangle_{\rho_\mathrm{st}^{(0)}}=0$.  The first contribution to $\langle J_\mathrm{meas} \rangle$ then arises from $\rho_\mathrm{st}^{(1)}$, which is linear in $\tau$, and is given by (see Appendix~\ref{app:perturbation}, Eq. \ref{eq:Danynotlike})
\begin{equation}
\braket{J_\mathrm{meas}}_{\rho_\mathrm{st}^{(1)}}  =\mathrm{tr}\left(-i[H,\rho_{\mathrm{st}}^{(0)}]J_\mathrm{meas}\right)\tau\,,
\end{equation}
up to corrections of order $\gamma_0/||H||$ which can be neglected for weak dissipation.
The measurement current operator, Eq.~\eqref{eq:measurement_current_definition}, has an overall factor of $1/\tau$ in its definition, making this contribution of the same order as the Hamiltonian current at zeroth order, $\braket{J_\mathrm{H}}_{\rho_\mathrm{st}^{(0)}}=\mathrm{tr}\left(\rho_{\mathrm{st}}^{(0)}J_\mathrm{H}\right)$.  For a generic observable we find that both currents are non-zero individually.  However, their sum cancels
\begin{equation}
\braket{J_\mathrm{meas}}_{\rho_\mathrm{st}^{(1)}}  + \braket{J_\mathrm{H}}_{\rho_\mathrm{st}^{(0)}}=0\,,
\end{equation}
up to corrections of order $\gamma_0/||H||$, as proved in Appendix~\ref{app:cancellation}.  This follows intuitively since we perform measurements that are local and non-overlapping in space.  Then, in the strict Zeno limit, particles cannot leave the domain over which the local measurements act, e.g. one of the unit cells in the Rice-Mele model, and therefore the total current must be zero.  This cancellation is born out by numerical results in the Rice-Mele model, as seen in Fig.~\ref{fig:current_limits}. This cancellation is discussed in more detail in Appendix \ref{app:zeno_limit_examples}.

\begin{figure}
\includegraphics[width=.45\textwidth]{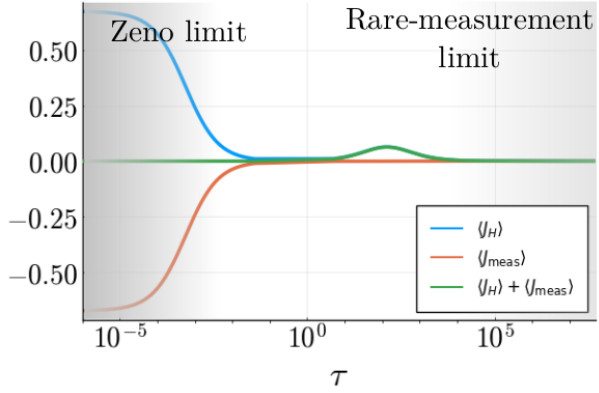}
    \caption{Hamiltonian, measurement, and total current dependence on measurement time $\tau$. The $\tau$-axis is logarithmic. The parameters of the plot are $t_1=1.5$, $t_2=1.0$, $V=3.0$, $\gamma_0=0.01$, $T=0.1$, $N=3$, and the measured observable is specified by $m_x=m_y=-m_z$. When $\tau$ is very small, we are in the Zeno limit, and the measurement current compensates for the Hamiltonian current.} 
    \label{fig:current_limits}
\end{figure}

\subsubsection{Local current loops in the Zeno limit}

\begin{figure}
\centering
\begin{overpic}[width=0.48\textwidth]{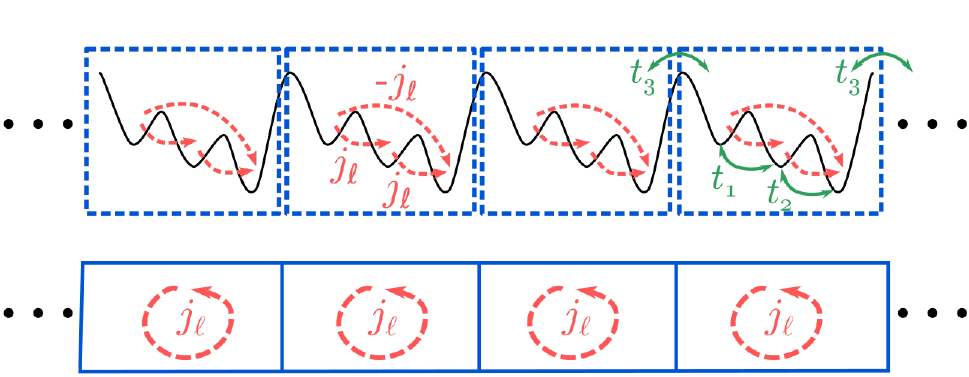}
\put(2, 35){(a)}
\put(2, 13){(b)}
\end{overpic}
\caption{Illustration of the three-site model. The measurements of an observable are performed on three-site unit cells, shown as dashed boxes in panel (a). Particles hop between sites with amplitudes $t_{1,2,3}$. In the Zeno limit, measurements prohibit charge transfer between the unit cells. This implies that the net current vanishes. However, as illustrated in (b), this does not prohibit the existence of loop currents within the unit cell (drawn in red).}
    \label{fig:loops}
\end{figure}

Thus far we have focused on dimerized Hamiltonians with bond measurements.  It is also interesting to consider Hamiltonians with a three-site unit cell acted on by three-site measurements.  Here we consider,
\begin{align}\label{eq:loop_hamiltonian}
    H &= \sum_{n=1}^N\Bigg[\Big(-t_1c_{3n-2}^{\dagger}c_{3n-1}-t_2c_{3n-1}^{\dagger}c_{3n}-t_3c_{3n}^{\dagger}c_{3n+1}\nonumber\\&\ \ \ \ \ \ \ \ \ \ +\mathrm{h.c.}\Big)+\frac{V}{2}\left(c_{3n-2}^{\dagger}c_{3n-2}-c_{3n}^{\dagger}c_{3n}\right)\Bigg]\,.
\end{align}
The model is pictorially represented in Fig.~\ref{fig:loops}(a). Each unit cell contains two bonds with hopping integrals $t_1$ and $t_2$ and the bond connecting two cells has hopping integral $t_3$. We choose the measured observable as a non-degenerate operator whose eigenstates are restricted within the unit cell. Following the argument for the Rice-Mele model, the net charge transfer has to vanish in the Zeno limit. However, this does not prohibit current loops within each unit cell, as shown in Fig.~\ref{fig:loops}(b). 

To show that loop currents can emerge we choose, for concreteness, a measurement operator whose eigenstates within the unit cell are,
\begin{align}
    \ket{\Psi_1(\alpha)}&=\frac{1}{\sqrt{3}}\left(\ket{1}+e^{i\alpha}\ket{2}+\ket{3}\right),\\
    \ket{\Psi_2(\alpha)} &=\frac{1}{\sqrt{2}}\left(\ket{1}-e^{i\alpha}\ket{2}\right), \\
    \ket{\Psi_3(\alpha)} &=\frac{1}{\sqrt{6}}\left(\ket{1}+e^{i\alpha}\ket{2}-2\ket{3}\right).
\end{align}
If the parameter $\alpha$ is not an integer multiple of $\pi$, complex phases are present. Then, the measured observable is not even under time reversal, and currents are not prohibited by symmetry. 

The steady-state currents are computed in Appendix \ref{app:zeno_limit_examples}.  There, it is found that the currents between sites satisfy $j^{1\to 2}=j^{2\to 3}=-j^{1\to 3}=j_\ell$, where
\begin{equation}
\begin{split}
    j_\ell &= \frac{1}{54}\Big[-\left(t_1+2t_2\right)\left(3u^{(0)}(\alpha)-2\right)\\
    &+3\left(t_1-t_2\right)\Delta u^{(0)}(\alpha)\Big]\sin(\alpha)\,.
\end{split}
\end{equation}
Here $u^{(0)}$ and $\Delta u^{(0)}$ are population parameters determined by solving the balance equation, Eq.~\eqref{eq:balance_equation}, and which depend on $\alpha$. Hence, the integrated total current vanishes for any value of $\alpha$, $j^{1\to 2}+j^{2\to 3}+2 j^{1\to 3}=0$, as expected. However, the nonvanishing of $j_\ell$ signals the presence of loop currents in the Zeno limit, as advertised.

\section{Floquet Measurements}\label{sec:Periodic}

We now consider a Floquet measurement scheme, in which measurements are performed periodically in time with a fixed period $\tau$~\cite{grunbaum2013, yin2024}.  In contrast to the Poisson measurement scheme, the system now lacks a true steady state.  Instead, the steady state is stroboscopic, with a time-periodic density matrix with period  $\tau$. As we now show, this allows Floquet measurements to probe the dynamics of the current following a measurement.  

\begin{figure}
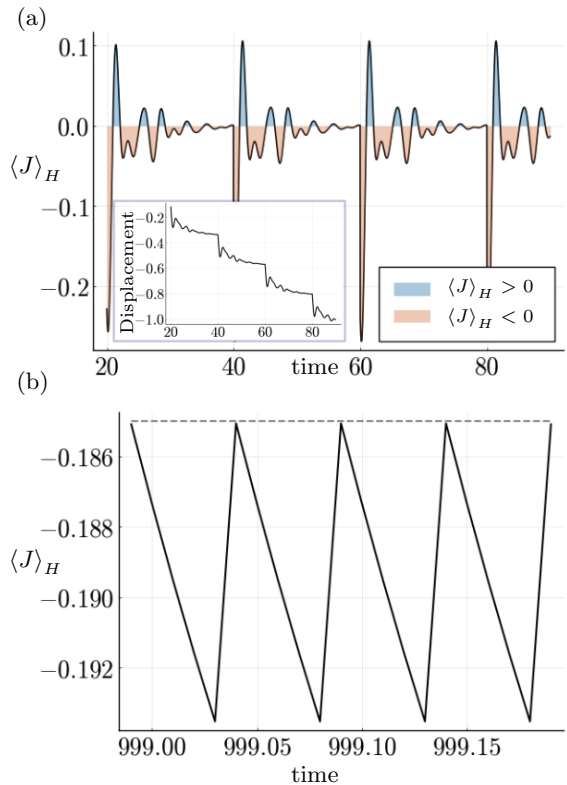

    \centering
    \begin{overpic}[width=0.4\textwidth]{images/current_dynamics_with_dissipation.pdf}
    \put(-2, 100){(a)}
    \put(-2, 51){(b)}
    \put(35, -2){time}
    \put(35, 53){time}
    \put(-3, 27){$\braket{J}_H$}
    \put(-3, 80){$\braket{J}_H$}
    \put(56, 64){\scriptsize{$\braket{J}_H>0$}}
    \put(56, 60){\scriptsize{$\braket{J}_H<0$}}
    \put(12.0, 58.4){\rotatebox{90}{\scriptsize{Displacement}}}
    \end{overpic}
    \caption{Expectation value of the Hamiltonian current as a function of time in the Floquet measurement scheme. The measured observable is given by $\munit{y}=\munit{z}=1/\sqrt{2}$, and $\munit{x}=0$. The parameters of the plot are $t_1=1.0$, $t_2=0.5$, $V=1.5$, $T=0.2$, $N=4$, and $\gamma_0=0.5$. In panel (a), the time between measurements is $\tau=20$, which is large compared to the relaxation time of the system and the system has enough time to relax until close to thermal equilibrium between measurements. In the inset, we show that for the given parameters, the particle moves to the left on average by time-integrating the Hamiltonian current and adding the measurement displacement. In b), $\tau=0.05$ and the evolution time dependence of the Hamiltonian current can be well approximated by taking low orders of expansion in $\tau$.  Near the Zeno limit the measurement charge displacement, not shown, is roughly compensated by the Hamiltonian current.}
    \label{fig:time_evolution}
\end{figure}

Figure \ref{fig:time_evolution} shows the total current for Floquet measurements of an inversion-preserving and time-reversal breaking observable in a system with broken inversion symmetry ($V\ne 0$).   The results were calculated by numerically evolving the density matrix and evaluating the expectation value of the Hamiltonian current once the system reaches a stroboscopic steady state.  Panel (a) shows the current in the rare-measurement limit $\tau_{\rm m}\gg 1/\gamma_0$.  In this limit, the system has time to relax to equilibrium between measurements. Then, the evolution of the current created by the measurement is revealed to oscillate on a time scale controlled by the Hamiltonian parameters and to decay with a rate of order $\gamma_0$.  There is a net charge transfer which we show by integrating the current over time, as seen in the inset of the figure.  This indicates a non-zero average current, $\bar{J}_H$, defined as the integrated charge transfer divided by the measurement period $\tau$,
\begin{align}
    \bar{J}_H=\frac{1}{\tau}\int_0^\tau dt\, \mathrm{tr}[\rho(t) J_H]
\end{align}
Panel (b) shows the Hamiltonian current near the Zeno limit.  Similar to the Poisson case, in the Zeno limit, the Hamiltonian current is compensated by the measurement charge displacement to give net zero displacement.

\begin{figure}
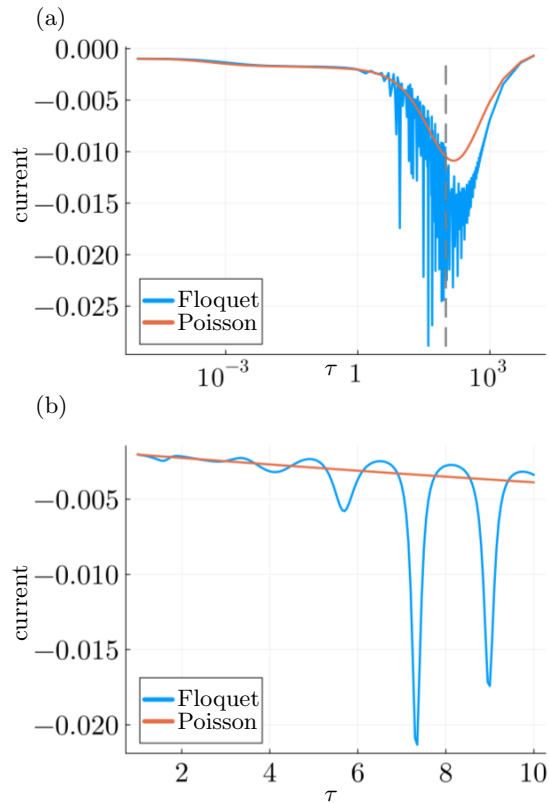

    \centering
    \begin{overpic}[width=0.4\textwidth]{images/tr_comparison.pdf}
    \put(-1, 20){\rotatebox{90}{current}}
    \put(-1, 75){\rotatebox{90}{current}}
    \put(2, 101){(a)}
    \put(2, 50){(b)}
    \put(40, -1){$\tau$}
    \put(40, 55){$\tau$}
    \put(20.7, 8.5){Poisson}
    \put(20.7, 11.5){Floquet}
    \put(20.7, 61.0){Poisson}
    \put(20.7, 64.0){Floquet}
    \end{overpic}
    \caption{Dependence of current on measurement time scale $\tau$. Floquet and Poisson measurement schemes are compared. The plot parameters are $t_1=1.0,\ t_2=0.5,\ V=3.0,\ T=0.1,\ N=3$, and $\gamma_0=0.01$. For Floquet measurements, $\bar{J}_H$ is shown. Measurements are of $\munit{y}=\munit{z}=1/\sqrt{2}$, and $\munit{x}=0$. Note that the Floquet measurement scheme displays resonances when the measurement period matches some of the internal system frequencies, as shown in (b), which focuses on a small window of panel (a) but on a linear scale.}
    \label{fig:periodic_vs_random}
\end{figure}

Figure \ref{fig:periodic_vs_random}(a) shows the current as a function of $\tau$ and compares it to the Poisson measurement protocol.  In both the Zeno and rare measurement limits, the two protocols converge to each other. Both schemes display a broad peak at $\tau\approx 1/\gamma_0$. However, the Floquet measurement scheme reveals a series of narrow resonances which, as shown in Fig.~\ref{fig:periodic_vs_random}(b) are almost periodic in $\tau$.  The origin of these resonances is the current oscillations induced by measurements, as seen in Fig.~\ref{fig:time_evolution}.  This suggests that Floquet measurements might be used as a method for probing the dynamics of quantum systems.

\section{Discussion}
Our paper provides a framework for analyzing currents in monitored quantum systems.  We showed that, beyond the usual Hamiltonian current, there is a measurement current, which can be comparable in magnitude.  The central topic has been the interplay of the symmetries of the Hamiltonian and the measurement observable, which dictates whether currents can exist, and their magnitude.  In particular, currents can be generated whether the measurement operators are time-reversal symmetric or not, due to the irreversibility of the system.  However, time-reversal breaking observables give rise to enhanced currents. It would be interesting to investigate to what extent these differences based on symmetry survive in more general systems, such as ones including interactions for example.  Similar considerations about symmetry may influence other properties of monitored quantum systems, beyond currents.  In particular, monitoring a particle in this way could be an interesting way of generating quantum active matter~\cite{adachi2022, zheng2023, khasseh2023, yamagishi2023}.

\begin{acknowledgments}

This work was supported by the Israel Science Foundation (grants No.~2038/21 , 2541/22, 1939/18) and by the NSF/BSF (2022605).

\end{acknowledgments}

\appendix

\section{Non-degenerate $T$-even observables}
\label{app:tEven}

In Section \ref{sec:timeReversalBreaking} we showed that a measurement operator $A$ that is either even or odd under the action of time reversal ($TAT^{-1}=\pm A$) does not give rise to DC currents, provided the measurement is done on a state that is time-reversal symmetric.  In this Appendix we show that, in the special case that $A$ is non-degenerate and time-reversal even, $TAT^{-1}=A$, then there are no DC currents after the measurement, regardless of the initial state.

When these conditions are satisfied, then each measurement induces a state that is time-reversal invariant, thereby precluding the occurrence of DC currents.   To see this observe that, in this scenario, the eigenstates of $A$ are also eigenstates of $T$, $T\ket{a}=\ket{a}$. These eigenstates therefore have no current initially, and as in Sec. ~\ref{sec:timeReversalBreaking}, they also do not carry any DC currents because they have balanced occupation numbers at $k$ and $-k$.  Thus even if the system starts out in a state that is carrying a current, the measurement switches off the current by forcing the system into a time-reversal symmetric state.

\section{Derivation of the measurement Lindbladian}
\label{app:measurementLindbladian}

We derive the Lindbladian for a system that undergoes repeated measurements of an observable $A$ at random and uncorrelated times with a rate $1/\tau$.

We begin by discretizing time into short intervals $\Delta t$ and consider how the density matrix changes between $t_n$ and $t_{n+1}=t_n+\Delta t$. 
In the small time $\Delta t$ the density matrix evolves with the Lindbladian $\mathcal{L}_\mathrm{HD}$ [defined in Eq.~\eqref{eq:lindbladian0}] into $e^{\mathcal{L}_\mathrm{HD}\Delta t}[\rho(t_n)]$.  At the end of the interval it is measured with probability $\frac{\Delta t}{\tau}$ causing it to change as in  \eqref{eq:density_matrix_mapping}, or not measured, with probability $1-\frac{\Delta t}{\tau}$. (The measurement may be regarded as occurring at the end of the interval since the order in which the measurement and the rest of the evolution occurs does not matter for a short interval of time, since they both involve infinitesimal changes to $\rho$ so they commute to first order.)
The density matrix at  $t_{n+1}=t_n+\Delta t$ is
\begin{equation}
\begin{split}
    \rho(t_{n+1})&=\left(1-\frac{\Delta t}{\tau}\right)e^{\mathcal{L}_\mathrm{HD}\Delta t}[\rho(t_{n})]\\ &+\frac{\Delta t}{\tau}\mathcal{P}_A\left[e^{\mathcal{L}_\mathrm{HD}\Delta t}[\rho(t_{n})]\right]+\mathcal{O}(\Delta t^2).
\end{split}
\end{equation}
In the limit $\Delta t\rightarrow 0$ one finds,
\begin{equation}
    \frac{d\rho(t)}{dt}=\mathcal{L}_\mathrm{HD}[\rho(t)]+\frac{1}{\tau}\left[\mathcal{P}_A[\rho(t)]-\rho(t)\right].
\end{equation}
This equation describes the evolution of the density matrix averaged over all measurement histories.  It retains the canonical Lindblad form with additional jump operators that are projectors $P_a$ onto eigenstates of the measured observable $A$.
When the measured observable acts on a two-site unit cell, the projectors for the specific cell can be written as $P_{\pm}=\frac{1}{2}\left(1\pm U\sigma_{\mathrm{z}}U^{\dagger}\right)$. The term in the measurement part of the Linbladian for this cell is then $\frac{1}{2\tau}\left(U\sigma_{\mathrm{z}}U^{\dagger}\rho(t)U\sigma_{\mathrm{z}}U^{\dagger}-\rho(t)\right)$. This has the form of a decohering bath with the decoherence rate $1/2\tau$. In this sense, the measurement apparatus is an engineered bath that picks a preferred basis and eliminates the coherences within it. 

\section{Perturbation theory for the steady state}
\label{app:perturbation}

In this appendix we develop a perturbation theory to compute the steady state for the Lindbladian
\begin{align}
    \calL=\calL_0+\calV \label{eq:lindbladperturbation}
\end{align}
where $\calV$ is a small perturbation relative to $\calL_0$.    The perturbation theory presented here is an application to Lindbladians of Brillouin-Wigner perturbation theory of quantum mechanics~\cite{lennard1930, brillouin1933, wigner1935}.  The reason for using this approach rather than more straightforward perturbation theory is that the unperturbed $\calL_0$ that we will choose has more than one steady state.  The steady state of the Lindbladian $\calL$, which is unique, will be obtained approximately from an effective Lindbladian acting within the space of steady states of $\calL_0$, that lifts the degeneracy.

The choice of $\calL_0$ and $\calV$ depends on the regime under consideration. The full Lindbladian in our system is
\begin{align}
\calL=\calH+\calL_\mathrm{m}+\calD\,.
\end{align}
Here $\calH[\rho]=-i\left[H,\rho\right]$ generates the unitary evolution, $\calD$ is the coupling to the bath, and $\calL_\mathrm{m}$ is the measurement, $\calL_\mathrm{m}=-\frac{1}{\tau}\calQ_A$, where $\calQ_A[\rho]=\rho-\sum_a P_a\rho P_a$ is a projector onto density matrices that have vanishing diagonal elements in the basis of the measurement observable $A$.  We denote the characteristic scale of the Hamiltonian by $||H||$, which is also the scale of $\calH$.  The scale of $\calD$ is the dissipation rate $\gamma_0$, which is assumed to be weak, $\gamma_0\ll ||H||$.  In the limit of fast measurements ($\frac1\tau\gg ||H||$) we take $\calL_0=-\frac{1}{\tau}\calQ_A$ and $\calV=\calH+\calD$. 
 For slow measurements ($\tau||H||\gg 1$) we take $\calL_0=\calH$ and $\calV=-\frac{1}{\tau}\calQ_A+\calD$.

To develop the perturbation theory, it is useful to think about $\calL$ as a superoperator that acts on the vector space of operators, including the density matrix $\rho$.
The steady-state condition is then written as $\calL\rho_\mathrm{st}=0$.  Define $\calP$ as a projector that fixes any steady state of $\calL_0$, while taking any eigenvector of $\calL_0$ with a nonzero eigenvalue to 0, so that 
\begin{align}
\calP\calL_0=\calL_0\calP=0\label{eq:LP0}    
\end{align}
Let $\calQ=1-\calP$ be the complementary projector.
We will split the steady state of the full Lindbladian into components in the two complementary spaces, $\rho_\mathrm{st}=\rho_P+\rho_Q$, where $\rho_P=\calP\rho_\mathrm{st}$ and $\rho_Q=\calQ\rho_\mathrm{st}$.  Then, the steady state condition is
\begin{align}
\label{eq:Lrho}
0=\calL\rho_\mathrm{st}=(\calL_0+\calV)(\rho_P+\rho_Q)=\calV\rho_P+(\calL_0+\calV)\rho_Q
\end{align}
using Eq.~\eqref{eq:LP0}.
This can be divided into two conditions by projecting it with $\calP$ and $\calQ$.
The latter gives 
\begin{align}
    \calQ(\calL_0+\calV)\calQ\rho_Q=-\calQ\calV\rho_P
\end{align}
and can be used to determine $\rho_Q$ in terms of $\rho_P$:
\begin{align}
\label{eq:rhoQ}
\rho_Q=-\calQ[\calQ(\calL_0+\calV)\calQ]^{-1}\calQ\calV\rho_P\,.
\end{align} 
While $\calQ[\calL_0+\calV]\calQ$ is non-invertible, as it has the same kernel as $\calQ$, the notation $\calQ[\calQ(\calL_0+\calV)\calQ]^{-1}\calQ$ indicates that the inverse is computed only within the complementary space.
Projecting Eq.~\eqref{eq:Lrho} with $\calP$ gives $\calP\calV\rho_P+\calP\calV\rho_Q=0$ which, combined with Eq.~\eqref{eq:rhoQ} becomes
\begin{align}
\calP\calV\rho_P-\calP\calV\calQ[\calQ(\calL_0+\calV)\calQ]^{-1}\calQ\calV\rho_P=0
\end{align}
Hence, $\rho_P$ is a steady state of the effective Lindbladian
\begin{align}
\calL_\mathrm{eff}=\calP\calV\calP-\calP\calV\calQ[\calQ(\calL_0+\calV)\calQ]^{-1}\calQ\calV\calP\,.
\end{align}
Once $\rho_P$ is known (based on this effective Lindbladian), $\rho_Q$ can be obtained from it using Eq.~\eqref{eq:rhoQ} to obtain $\rho_{\mathrm{st}}=\rho_P+\rho_Q$.  To second order in $\calV$, the effective Lindblad operator is
\begin{align}
    \calL_\mathrm{eff}\approx \calP\calV\calP-\calP\calV\calQ (\calL_0)^{-1}\calQ\calV\calP.
    \label{eq:effectiveL}
\end{align}
Perturbation theory is justified provided that there is a gap from the zero eigenvalues of $\calL_0$ to the nonzero eigenvalues.

\begin{figure}
    \centering
    \includegraphics[width=0.48\textwidth]{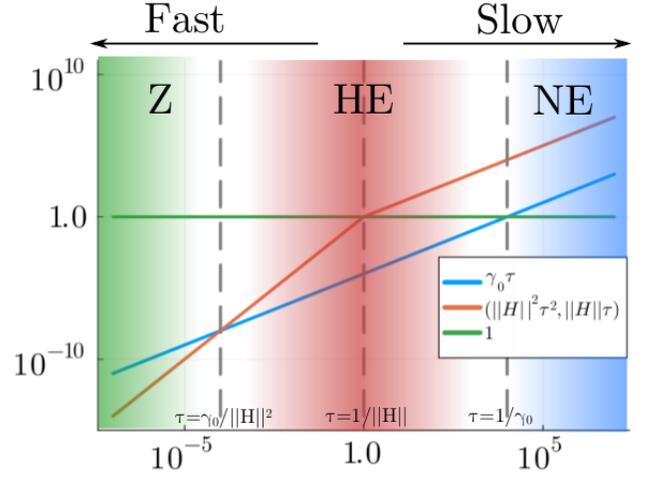}
    \caption{Schematic depiction of the different dynamical regimes.  Different dynamical scales are shown as a function of measurement time $\tau$. They are: the Hamiltonian scale ($||H||\tau$ or $||H||^2\tau^2$ depending on whether measurements are fast or slow), the dissipative scale ($\gamma_0\tau$), and the measurement scale (constant in $\tau$), all normalized by the measurement rate.  Both axes are logarithmic. In the Zeno (Z) regime, the Hamiltonian dynamics is negligible and the steady state is determined by the interplay of measurements and dissipation. In the near-equilibrium (NE) regime, the dissipation has enough time to equilibrate the system between measurements.   In the high entropy (HE) regime, the dissipation is the weakest scale and the system approaches an infinite temperature state. The parameters in the plot are $||H||=1$, $\gamma_0=10^{-4}$.}
    \label{fig:scales}
\end{figure}

We will next apply this formalism in the limits of fast and slow measurements, in turn.   We will find various dynamical regimes as the measurement time is varied, which are summarized in Fig.~\ref{fig:scales}.

\subsection{Fast measurement limit $||H||\tau\ll 1$}

When measurements are fast relative to the Hamiltonian timescale, $||H||\tau\ll 1$, we take $\calL_0=-\frac{1}{\tau}\calQ_A$ as the dominant term in the Lindbladian, and $\calV=\calH+\calD$ as the perturbation.  In this case, the projector onto the steady states of $\calL_0$ is $\calP_A=1-\calQ_A$, whose action on $\rho$ is $\calP_A[\rho]=\sum_a P_a \rho P_a$. This is highly degenerate, as any density matrix that is diagonal in the measurement basis,
\begin{align}
\rho=\sum_a w_a\ket{a}\bra{a},
\end{align}
is a steady state of $\calL_0$, regardless of the value of the occupations $w_a$. The perturbation $\calV$ lifts this degeneracy. Equation \eqref{eq:effectiveL} gives
\begin{align}
\calL_\mathrm{eff}=\calP_A \calV \calP_A+\tau \calP_A \calV \calQ_A \calV \calP_A\,.
\end{align}
The unitary evolution vanishes at leading order, $\calP_A\calH\calP_A=0$, as can be seen by acting on a general density matrix $\rho$,
\begin{align}
\calP_A\calH\calP_A[\rho]&=\sum_a P_a\left(-i\left[H,\sum_b P_b \rho P_b\right]\right)P_a\\
&=-i\sum_a\left[P_a H P_a,P_a\rho P_a\right]=0\,,
\end{align}
which vanishes for a non-degenerate observable $A$.
Therefore, $\calH$ contributes to $\calL_\mathrm{eff}$ starting at second order.  To leading non-vanishing order in $\gamma_0$ and $||H||$,
\begin{align}
\calL_\mathrm{eff}=\calP_A \left(\calD +\tau \calH^2\right) \calP_A\,.\label{eq:effectivefastmeasurements}
\end{align}
The condition $\calL_\mathrm{eff}\rho_P=0$ next fixes the steady-state occupations $w_a$ in the measurement basis, $\rho_P$.  The off-diagonal elements of $\rho_\mathrm{st}$ in the measurement basis, $\rho_Q$, are then given by Eq.~\eqref{eq:rhoQ}.  They are subdominant in $\tau\calV$ relative to the diagonal elements.

The form of $\rho_{\mathrm{st}}$ depends on which of the terms in Eq.~\eqref{eq:effectivefastmeasurements} is larger. There is thus a crossover as a function of $\tau$ at $\tau_{\mathrm{Z}}\equiv \gamma_0/||H||^2$.  For ultrafast measurements, $\tau\ll \tau_{\mathrm{Z}}$, the steady state occupations are determined by the condition $\calP_A\calD\calP_A\rho_{\mathrm{st}}^{(0)}=0$, which yields the balance equation
\begin{equation}\label{eq:global_balance equation}
    \sum_{a'\neq a}w_{a'}^{(0)}T^{a'\rightarrow a}=w_a^{(0)}\sum_{a'\neq a}T^{a\rightarrow a'},
\end{equation}
where $T^{a\rightarrow a'}=\sum_\alpha |\braket{a'|L_\alpha|a}|^2$ is the transition rate from state $a$ to $a'$.  The subscript $(0)$ indicates that these are the occupations to zeroth order in $\tau$.  The von Neumann entropy in this case saturates to a value that is independent of both $\gamma_0$ and $\tau$.  This is what we refer to as the Zeno limit.

On the other hand, for $\tau_{\mathrm{Z}}\ll \tau \ll 1/||H||$, the Hamiltonian plays a dominant role over the dissipation in Eq.~\eqref{eq:effectivefastmeasurements}.  If the dissipation is fully neglected, then according to the arguments in Sec.~\ref{sec:thermalBath}, the steady state will be an infinite temperature state, $\rho_\infty=\frac{\mathbb{1}}{{\mathcal{N}}}$. Adding the dissipation $\calD$ prevents the system from reaching this state.  Let
\begin{equation}
\rho_P=\rho_\infty+\delta\rho_P,
\end{equation}
where $\delta\rho_P$ is proportional to $\gamma_0$.
The correction can be found by solving $\calL_\mathrm{eff}\rho_P=0$ to first order in $\gamma_0$ for $\delta\rho_P$.
This gives
\begin{align}
\delta\rho_P&=-\frac{1}{\tau}\calP_A(\calP_A\calH^2\calP_A)^{-1} \calP_A\calD \rho_\infty\,,
\end{align}
which is of order $\gamma_0/(\tau ||H||^2)$. 
The complementary part of the density matrix, $\rho_Q=\calQ_A\rho_\mathrm{st}$, is found to be smaller:
\begin{align}
\rho_Q&=\tau\calQ_A(\calH+\calD)(\rho_\infty+\delta\rho_P)\\
&\approx \calQ_A\calD\tau\rho_\infty-\calQ_A\calH\calP_A(\calP_A\calH^2\calP_A)^{-1}\calP_A\calD\rho_\infty.\nonumber
\end{align}
These terms are smaller than $\delta\rho_P$ by factors of $(\tau||H||)^2$ and $(\tau||H||)$ respectively, so they can be neglected in the fast measurement limit. 

We turn our attention to the von Neumann entropy,
\begin{equation}
\begin{split}
    &S[\rho_{\mathrm{st}}]=-\mathrm{tr}\left[\left(\rho_\infty+\delta\rho\right)\log\left(\rho_\infty+\delta\rho\right)\right]
\end{split}
\end{equation}
Expanding $\log(\rho_\infty+\delta\rho)=\log((\mathbb{1}+\mathcal{N}\delta\rho)/\mathcal{N})=-\log\mathcal{N}+\mathcal{N}\delta\rho-\frac{\mathcal{N}^2}{2}\delta\rho^2+\mathcal{O}\left(\gamma_0^2\right)$
then, since $\delta\rho$ is traceless, the entropy is to order $\gamma_0^2$
\begin{equation}
    S[\rho_{\mathrm{st}}]=\log\mathcal{N}-\frac{\mathcal{N}}{2}\mathrm{tr}\left[\delta\rho^2\right]\,.
\end{equation}
The relative entropy as defined in Eq.~\eqref{eq:relative_entropy} is then
\begin{equation}\label{eq:relative_entropy_trace}
    \Delta s=\frac{\mathcal{N}}{2\log\left(\mathcal{N}\right)}\mathrm{tr}\left[\delta\rho^2\right]\,.
\end{equation}
Since $\delta\rho$ is of order $\gamma_0/(\tau ||H||^2)$, $\Delta s$ scales with $\tau$ as 
\begin{align}
\label{eq:deltaS}
\Delta s\propto \frac{\gamma_0^2}{||H||^4 \tau^2}  
\end{align}
in the regime $\gamma_0/||H||^2\ll \tau\ll 1/||H||$, explaining the scaling seen numerically in Sec. \ref{sec:example}.

In Appendix~\ref{app:cancellation} we compute currents in the Zeno limit. The measurement current is non-zero only in the complement of the measurement subspace, 
\begin{align}
\braket{J_{\mathrm{meas}}}_{\mathcal{P}_A\left[\rho\right]}=0\,.\label{eq:JmeasNoP}
\end{align}
Therefore, we need the off-diagonal entries of the density matrix when $\tau\ll\frac1\gamma_0$. Using Eq.~\eqref{eq:rhoQ}, we find
\begin{align}
\calQ_A\rho^{(1)}_\mathrm{st}&=\tau\calQ_A\calH\calP_A\rho_\mathrm{st}^{(0)}\nonumber\\
&=-i\tau[H,\rho_{\mathrm{st}}^{(0)}].
\label{eq:Danynotlike}
\end{align}
where we have neglected terms of order $\tau \gamma_0$, and where the superscript $(1)$ indicates linear order in $\tau$. In the second line we used $\calQ_A=1-\calP_A$, $\calP_A \calH \calP_A=0$ and the expression for $\calH$ as the commutator with $H$. 

\subsection{Slow measurement limit $\tau||H||\gg 1$}

In the opposite limit, when measurements are slow relative to the Hamiltonian timescale, we take $\calL_0=\calH$ and $\calV=\calL_\mathrm{m}+\calD=-\frac{1}{\tau}\calQ_A+\calD$.  Any diagonal density matrix in the Hamiltonian basis is a steady state of $\calL_0$.  The projector onto the space of steady states is $\calP_H$, whose action on density matrices is $\calP_H[\rho]=\sum_{E} P_E \rho P_E$.  Here, $P_E$ is a projector into the subspace of states with energy $E$. Equation~\eqref{eq:effectiveL} then gives
\begin{align}
\label{eq:effLslow}
\calL_\mathrm{eff}=\calP_H\left(-\frac{1}{\tau}\calQ_A+\calD\right)\calP_H
\end{align}
where the terms above are nonvanishing and hence the leading order contributions in $\calV$.  The condition $\calL_\mathrm{eff}\rho_\mathrm{st}=0$ fixes the steady-state occupations in the Hamiltonian basis. There is a crossover as a function of $\tau$ at $\tau_\mathrm{R}\equiv 1/\gamma_0$.  For $\tau\gg \tau_\mathrm{R}$, the steady state occupations are given by $\calP_H\calD\calP_H\rho_\mathrm{st}=0,$ which yields the equilibrium state at the temperature of the bath.

On the other hand, for $1/||H||\ll \tau\ll \tau_\mathrm{R},$ $\calL_\mathrm{eff}\approx-\frac{1}{\tau}\calP_H\calQ_A\calP_H$, whose steady state is the infinite temperature state (since it is simultaneously diagonal in the Hamiltonian and measurement bases).  The dissipation $\calD$ prevents the system from reaching infinite temperature.  Writing $\rho_{\mathrm{st}}=\rho_\infty+\delta\rho$ for the steady state of Eq.~\eqref{eq:effLslow}, we find that $\delta\rho$ is of order $\gamma_0\tau$.  Plugging this into Eq.~\eqref{eq:deltaS} yields
\begin{align}
\Delta s\propto \gamma_0^2\tau^2
\end{align}
which is the scaling of the relative entropy with $\tau$ in the regime $1/||H||\ll \tau\ll \tau_\mathrm{R}$.

\section{No charge transfer in the Zeno limit \label{app:cancellation}}

Using the results from Appendix~\ref{app:perturbation}, we show that the total current vanishes in the Zeno limit. By Eq. ~\eqref{eq:TotalJH}, the total Hamiltonian current is $-i[\hat{x},H]$ and its expectation value for the density matrix of the Zeno limit, Eq. ~\eqref{eq:zeroth_rho}, is
\begin{equation}
    \braket{J_H}_{\rho_{\mathrm{st}}^{(0)}}=i\sum_aw_a^{(0)}\mathrm{tr}\left(\left[H,\hat{x}\right]P_a\right).
\end{equation}

The measurement current is non-zero only in the complement of the measurement subspace, $\braket{J_{\mathrm{meas}}}_{\mathcal{P}_A\left[\rho\right]}=0$ (see Eq. \eqref{eq:TotalJm}) Therefore, we evaluate the expectation value with respect to $\rho_Q$. This part of the density matrix is smaller by a factor of $\tau||H||$.
However, its contribution to the total current is of the same order as the Hamiltonian current at leading order, since the measurement current has an overall factor of $1/\tau$ in its definition, see Eq.~\eqref{eq:measurement_current_definition}.   
By
 Eq.~\eqref{eq:Danynotlike} 
\begin{equation}\begin{split}
\mathcal{Q}_A\rho_{\mathrm{st}}^{(1)}&=-i\left[H, \rho_{\mathrm{st}}^{(0)}\right]\tau+\mathcal{O}\left(\gamma_0\right)\\
&=-i\sum_a\left[H, P_a\right]w_a^{(0)}\tau+\mathcal{O}\left(\gamma_0\right).
\end{split}
\end{equation} The expectation value of the integrated measurement current for this density matrix is
\begin{equation}
\braket{J_{\mathrm{meas}}}_{\rho_{\mathrm{st}}^{(1)}}=i\sum_{a, a'}w_a^{(0)}\mathrm{tr}\big(\left[H, P_a\right]
\left(\hat{x}-P_{a'}\hat{x}P_{a'}\right)\big).
\end{equation}
The first term cancels the Hamiltonian current since $\mathrm{tr}[H,P_a]\hat{x}=-\mathrm{tr}[H,\hat{x}]P_a$.  Thus, the current will vanish if the last term is zero. The last term is
\begin{equation}
\begin{split}
    &-i\sum_{a, a'}w_a^{(0)}\mathrm{tr}\left(\left[H, P_a\right]P_{a'}\hat{x}P_{a'}\right)\\
    &=-i\sum_{a,a'}w_a^{(0)}\mathrm{tr}\left(HP_aP_{a'}\hat{x}P_{a'}-P_aHP_{a'}\hat{x}P_{a'}\right)\\
    &=-i\sum_a w_a^{(0)}\mathrm{tr}\left(HP_a\hat{x}P_a-P_aHP_a\hat{x}P_a\right)=0,
\end{split}
\end{equation}
from the projector identity $P_aP_{a'}=\delta_{a, a'}P_a$ and the cyclic property of the trace. 

\section{Two-site and three-site measurements in the Zeno limit}\label{app:zeno_limit_examples}

We will now consider local currents in the Zeno limit.
In this limit, for the dynamics we consider [see Fig.~\ref{fig:measurement_protocol}], measurements destroy coherence between unit cells.  The lack of coherence between unit cells prevents Hamiltonian currents from developing between unit cells.  Since measurements do not connect unit cells, no measurement currents will develop either.  Therefore, in what follows, we study the dynamics within a unit cell.

To ease notation, given an operator $O$, we define the operator $O^{\mathrm{c}}$
\begin{equation}
    O^{\mathrm{c}}=N P_n O P_n,
\end{equation}
where $P_n$ is the projector onto a specific unit cell $n$. The factor $N$ takes into account the fact that the particle can be in any of the $N$ unit cells, and ensures proper normalization.
In particular, the density matrix in the Zeno limit is a sum over unit cells, so no information is lost by considering $\rho^\mathrm{c}$ for a specific cell.  Consequently, in the calculation of the currents, e.g., the Hamiltonian current $\langle J_H\rangle=i\ \mathrm{tr}\ \rho_{\mathrm{st}}[H,\hat{x}]$, the Hamiltonian may also be replaced by $H^c$.  

\subsection{Two-site measurements}

If the measurement subspace is two-dimensional, as in Sec.~\ref{sec:example}, the density matrix projected onto a unit cell can be written as
\begin{equation}
    \rho^{\mathrm{c}}_{\mathrm{st}}=\frac{1}{2}\left[1+\left(2w_\mathrm{L}^{(0)}-1\right)\mathbf{\hat{m}}\cdot \boldsymbol{\sigma}\right],
\end{equation}
where $w_\mathrm{L}^{(0)}$ is the occupation probability of the left site after the control $U$ is applied. It is determined by Eq.~\eqref{eq:global_balance equation},
\begin{equation}
    w_{\mathrm{L}}^{(0)}=\frac{1}{1+\frac{T^{1\to 2}}{T^{2\to 1}}},
\end{equation}
where $T^{1\to 2}=\sum_{\alpha}|\braket{2|UL_{\alpha}U^\dagger|1}|^2$ is the transition probability between state $U^\dagger\ket{1}$ and $U^\dagger\ket{2}$, where $\ket{1}$ and $\ket{2}$ are the position eigenstates. Consider a general form for the Hamiltonian projected into the two-site unit cell:
\begin{eqnarray}
    H^{\mathrm{c}} = h_0\sigma_0+\mathbf{h}\cdot\boldsymbol{\sigma},
\end{eqnarray}
where $\sigma_0$ is the identity matrix and the vector $\mathbf{h}$ specifies the subspace Hamiltonian. Then the operator for the Hamiltonian current, $i[H^c,\hat{x}]=i[H^c,-\frac{\sigma_z}{2}]$, can be written as
\begin{equation}
    J_H^{\mathrm{c}}=\left(\mathbf{h}\times\hat{z}\right)\cdot\boldsymbol{\sigma}.
\end{equation}

The expectation value of the Hamiltonian current is then
\begin{equation}
\begin{split}\braket{J_H^{\mathrm{c}}}&=\left(w_{\mathrm{L}}^{(0)}-\frac{1}{2}\right)\mathrm{tr}\left[\left(\mathbf{h}\times\hat{z}\right)\cdot\boldsymbol{\sigma}\left(\mathbf{\hat{m}}\cdot \boldsymbol{\sigma}\right)\right]\\
&=-2\left(w_{\mathrm{L}}^{(0)}-\frac{1}{2}\right)\left(\mathbf{h}\times\mathbf{\hat{m}}\right)_{\mathrm{z}}.
\end{split}
\end{equation}

The effective two-site measurement current operator can be found by inserting the projectors onto the measured states $P_\pm=\frac12(1\pm\mathbf{\hat{m}}\cdot\boldsymbol{\sigma})$ into the general measurement current expression (Eq.\eqref{eq:TotalJm}).  We find
\begin{equation}
J_{\mathrm{meas}}^{\mathrm{c}}=\frac{1}{2\tau}\left(\sigma_{\mathrm{z}}-m_z\left(\mathbf{\hat{m}}\cdot\boldsymbol{\sigma}\right)\right).
\end{equation}
The measurement current expectation value is zero within the measurement subspace, and the dominant contribution comes from $\mathcal{Q}_A\rho_{\mathrm{st}}^{(1)}$, (Eq. \eqref{eq:Danynotlike}). 
 By projecting to the measurement subspace we get that the expectation value of the measurement current is
\begin{equation}
\begin{split}
    &\braket{J_{\mathrm{meas}}^{\mathrm{c}}}=-\frac{i}{2}\mathrm{tr}\left[\left(\sigma_{\mathrm{z}}-m_z\left(\mathbf{\hat{m}}\cdot\boldsymbol{\sigma}\right)\right)\left(\left[H,\rho_{\mathrm{st}}^{(0)}\right]^{\mathrm{c}}\right)\right]\\
    &=\left(w_\mathrm{L}^{(0)}-\frac{1}{2}\right)\mathrm{tr}\left[\left(\sigma_{\mathrm{z}}-m_z\left(\mathbf{\hat{m}}\cdot\boldsymbol{\sigma}\right)\right)\left(\mathbf{h}\times\mathbf{\hat{m}}\right)\cdot\boldsymbol{\sigma}\right]\\
    &=2\left(w_\mathrm{L}^{(0)}-\frac{1}{2}\right)\left(\mathbf{h}\times\mathbf{\hat{m}}\right)_{z},
\end{split}
\end{equation}
where in the last line we used $\left(\mathbf{h}\times\mathbf{\hat{m}}\right)\cdot\mathbf{\hat{m}}=0$ and $\mathrm{tr}\left(\sigma_i\right)=0$. Note that the Hamiltonian current is exactly compensated by the measurement current.

\begin{figure}
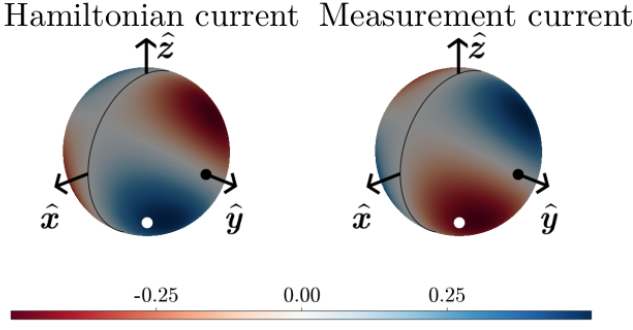

    \centering
    \begin{overpic}[width=.48\textwidth]{images/zeno_limit_spheres.pdf}
    \end{overpic}
    \caption{Hamiltonian and measurement currents are plotted on the Bloch sphere in the Zeno limit. The observable that is measured in Fig.~\ref{fig:current_limits} is marked by the white dot. In that figure, it can be seen that these two currents cancel out in the Zeno limit and that this cancellation doesn't persist to larger measurement times. The parameters are the same as in Fig.~\ref{fig:current_limits}.} 
    \label{fig:zeno_limit_spheres}
\end{figure}

Despite the cancellation, it is interesting to look at the Hamiltonian and measurement currents separately.  In the Zeno limit, the steady-state currents for the Rice-Mele model ($\mathbf{h}=-t_1 \hat{\mathbf{e}}_x+\frac{V}{2} \hat{\mathbf{e}}_z$) is
\begin{equation}
\braket{J_\mathrm{H}}_{\rho_\mathrm{st}^{(0)}}=-\braket{J_\mathrm{meas}}_{\rho_\mathrm{st}^{(1)}}=2t_1 \left(w_L^{(0)}-1/2\right)m_{\mathrm{y}}\,.
\end{equation} 
Figure~\ref{fig:zeno_limit_spheres} illustrates these currents.  Note that each current vanishes separately on the great circle $\munit{y}=0$, corresponding to $T$-even observables. It also vanishes for measurements of the bond current itself, $\munit{y}=\pm 1$, when $w_\mathrm{L}^{(0)}=\frac{1}{2}$.

\subsection{Three-site measurements and loop currents}
In analogy to the Rice-Mele Hamiltonian, we use a similar Hamiltonian, but with three sites in the unit cell instead of two,
\begin{align}\label{eq:three_site_hamiltonian}
    H =& \sum_{n=1}^N\Bigg[\Big(-t_1c_{3n-2}^{\dagger}c_{3n-1}-t_2c_{3n-1}^{\dagger}c_{3n}\\-&t_3c_{3n}^{\dagger}c_{3n+1}+\mathrm{h.c.}\Big)+\frac{V}{2}\left(c_{3n-2}^{\dagger}c_{3n-2}-c_{3n}^{\dagger}c_{3n}\right)\Bigg]\,.\nonumber
\end{align}
We perform measurements within the three-dimensional unit cell. The measurement subspace Hamiltonian can be written in the Gell-Mann basis as 
\begin{equation}
    H^{\mathrm{c}}= -t_1\lambda_1-t_2\lambda_6+\frac{V}{2}\left(\frac{1}{2}\lambda_3+\frac{\sqrt{3}}{2}\lambda_8\right),
\end{equation}
where $\lambda_i$-s are Gell-Mann matrices. Similarly we write the expression for the Hamiltonian current,
\begin{equation}
    J_H^{\mathrm{c}}=t_1\lambda_2+t_2\lambda_7.
\end{equation}

In order to obtain nonzero currents the measurements should act on the three sites of the unit cell.  This allows the density matrix to have coherences between all three sites in the unit cell, allowing for currents between all pairs of sites so that a circular current can form. Moreover, for the current to be non-zero, the measured eigenstates should not have a well-defined parity under time reversal, indicating that complex amplitudes must be involved. An example of a vector that satisfies these requirements is
\begin{equation}
    \ket{\Psi_1(\alpha)}=\frac{1}{\sqrt{3}}\left(\ket{1}+e^{i\alpha}\ket{2}+\ket{3}\right),
\end{equation}
where $\alpha$ is a phase that is not an integer multiple of $\pi$. We choose the two other eigenvectors to be
\begin{align}
    \ket{\Psi_2(\alpha)} &=\frac{1}{\sqrt{2}}\left(\ket{1}-e^{i\alpha}\ket{2}\right) \\
    \ket{\Psi_3(\alpha)} &=\frac{1}{\sqrt{6}}\left(\ket{1}+e^{i\alpha}\ket{2}-2\ket{3}\right).
\end{align}
Measurement of the non-degenerate observable $A^{\mathrm{c}}(\alpha)=\sum_{i=1}^{3}a_i\ket{\Psi_i(\alpha)}\bra{\Psi_i(\alpha)}$ where  each $a_i$ is unique will yield these eigenvectors. For instance, if $\alpha=\frac{\pi}{2}$ the projectors onto these three eigenstates, $P_i(\alpha)=\ket{\Psi_i(\alpha)}\bra{\Psi(\alpha)}$, are
\begin{align}
    P_1\left(\frac{\pi}{2}\right)  &=\frac{1}{3}\left(1+\lambda_4+\lambda_2-\lambda_7\right),\\
    P_2\left(\frac{\pi}{2}\right)  &=\frac{1}{3}\left(1+\frac{\sqrt{3}}{2}\lambda_8-\frac{3}{2}\lambda_2\right),\\
    P_3\left(\frac{\pi}{2}\right) &=\frac{1}{3}\left(1-\lambda_4-\frac{\sqrt{3}}{2}\lambda_8+\frac{\lambda_2}{2}+\lambda_7\right).
\end{align}

Using the discrete translational symmetry of the problem we write the steady-state density matrix as
\begin{equation}
\begin{split}
\rho^{\mathrm{c}}_{\mathrm{st}}\left(\alpha\right)=\frac{1}{3}\Big[1-&\frac{u^{(0)}+\Delta u^{(0)}}{2}P_1\left(\alpha\right)\\-&\frac{u^{(0)}-\Delta u^{(0)}}{2}P_2\left(\alpha\right)\\
+&u^{(0)}P_3\left(\alpha\right)\Big],
\end{split}
\end{equation}
where $u^{(0)}$ and $\Delta u^{(0)}$ are parameters that can be determined by solving the balance equation, Eq.~\eqref{eq:balance_equation}. 

The expectation value of the Hamiltonian current is
\begin{equation}\label{eq:net_hamiltonian_three}
\begin{split}
\braket{J_H^{\mathrm{c}}}\left(\alpha\right)=&\frac{1}{18}\Big[3\left(2 t_2+t_1\right) u^{(0)}\\
   -&\left(5t_1-2t_2\right)\Delta u^{(0)}\Big] \sin\left(\alpha\right).
\end{split}
\end{equation}
On the other hand, the expectation values of the measurement currents between any of three possible pairs of sites come out to be (after calculating the traces of the products of matrices in Eq.\eqref{eq:measurement_current_definition}):
\begin{align}
\braket{j_{\mathrm{meas}}^{\mathrm{c}}}_{1\rightarrow 2}\left(\alpha\right)=&\frac{1}{18}\Big[-2 \left(t_1-t_2\right) u^{(0)}\nonumber\\
   +&\left(4t_1+t_2\right)\Delta u^{(0)}\Big] \sin\left(\alpha\right),
\end{align}
\begin{align}
\braket{j_{\mathrm{meas}}^{\mathrm{c}}}_{2\rightarrow 3}\left(\alpha\right)=&\frac{1}{18}\Big[ \left(t_1-4t_2\right)u^{(0)}\nonumber\\
   -&\left(t_1+t_2\right)\Delta u^{(0)}\Big] \sin\left(\alpha\right),\\ \label{eq:zeno13}
\braket{j_{\mathrm{meas}}^{\mathrm{c}}}_{1\rightarrow 3}\left(\alpha\right)=&\frac{1}{18}\Big[ -\left(t_1+2t_2\right)u^{(0)} \nonumber\\
   +&\left(t_1-t_2\right)\Delta u^{(0)}\Big] \sin\left(\alpha\right).
\end{align}
If we calculate the total measurement current $\braket{j^\mathrm{c}_\mathrm{meas}}_{1\rightarrow 2}+\braket{j^\mathrm{c}_\mathrm{meas}}_{2\rightarrow 3}+2\braket{j^\mathrm{c}_\mathrm{meas}}_{1\rightarrow 3}$
we see that the charge transferred by measurement is exactly compensated by the charge transferred via the Hamiltonian current,
\begin{equation}
\braket{J^{\mathrm{c}}_{\mathrm{meas}}+J^{\mathrm{c}}_{\mathrm{H}}}=0.
\end{equation}
\\However, the current between pairs of sites does not vanish.  The Hamiltonian contains only nearest neighbor couplings, which means that $\braket{j_{H}^{\mathrm{c}}}_{1\rightarrow 3}$ is zero. This is evidently not the case for the measurement current between these two sites as seen in Eq.~\eqref{eq:zeno13}.  This indicates that a steady-state current flows in a loop that encompasses the three sites. In fact, if the Hamiltonian current is divided into currents between pairs of sites, one finds that $\braket{j^\mathrm{c}_\mathrm{meas}+j^\mathrm{c}_\mathrm{H}}_{1\rightarrow 2}=\braket{j^\mathrm{c}_\mathrm{meas}+j^\mathrm{c}_\mathrm{H}}_{2\rightarrow 3}=\braket{j^\mathrm{c}_\mathrm{meas}+j^\mathrm{c}_\mathrm{H}}_{3\rightarrow 1}\neq 0$.  This is seen using the expression for the Hamiltonian current $j_H^{x\rightarrow y}=i(P_xHP_y-P_yHP_x)$ between pairs of sites.

\bibliography{measurement_currents}
\end{document}